# Biomolecular crystallization in microfluidic devices

**Nadine CANDONI[1], Romain GROSSIER[1] and Stéphane VEESLER[1]**

[1]Centre Interdisciplinaire de Nanoscience de Marseille (CINaM), Aix-Marseille Université, Marseille, France

Corresponding author: Nadine CANDONI, nadine.candoni@univ-amu.fr

**Abstract**
This chapter presents an overview of microfluidic devices reported in the literature, used to develop methodologies for nucleation of biomolecules, with crystal size control, and for collecting thermodynamic and kinetic data. Part I is dedicated to the properties of microfluidic devices through materials used for their fabrication and for crystals analysis. Part II describes the variety of microfluidic devices available and how to handle them to produce flows, droplets and/or wells of micrometer size. These devices use crystallization methods inspired by batch processes and they are mainly used for protein crystallization. Part III focuses on fundamental properties of biomolecule crystallization determined using droplet-based microfluidics: nucleation kinetics, nucleation rate and effective interfacial energy crystal/solution. Part IV explains how the kinetic effect of confinement due to micrometer size, and so nanovolumes, leads to isolation of different phases. These latter are characterized by X-Ray Diffraction (XRD) and methods to minimize manual handling of crystals for XRD are also presented, with appropriate equipment to store the crystals.

**Table of content**

# Introduction

In the past decades, microfluidic tools to control and manipulate flows at sub-millimeter scale, have increasingly been used in a lab-on-chip approach (Leng and Salmon 2009). This has allowed the number of experiments to be increased and the amount of material to be decreased, allowing statistical approach for studying stochastic phenomena such as nucleation (Leng et al. 2009, Song et al. 2006, Shi et al. 2017) Crystallization is used for biomolecule separation, purification processes, control of product properties and analytical purposes (Erkamp et al. 2023). For instance, protein crystallization is a key step in the determination of the three-dimensional structure of proteins by X-ray diffraction, allowing a better understanding of their biological functions. It is also expected to replace the traditional downstream processing for protein-based biopharmaceuticals by protein crystallization due to its advantages in stability, storage, and delivery as pointed-out by Ferreira and Castro (2023). Microfluidics renders possible screening crystallization conditions and then optimizing one of them to obtain crystals with controlled size, habit or phase, notably for biomolecules that are expensive to produce and of limited availability. Furthermore, reducing the volume of crystallization also makes observation of few crystals at the micrometer scale possible. Thus, the resolution of detection is increased with optical microscope or spectroscopic techniques for fundamental studies or process control (Lambert et al. 2023). This chapter presents an overview of microfluidic devices reported in the literature, used to develop methodologies for nucleation of biomolecules, with crystal size control and for collecting thermodynamic and kinetic data. Part I is dedicated to the properties of microfluidic devices through materials used for their fabrication and for crystals analysis. Part II describes the variety of microfluidic devices available, that use crystallization methods inspired by batch processes. Part III focuses on fundamental properties of biomolecule crystallization determined using droplet-based microfluidics: nucleation kinetics, nucleation rate and effective interfacial energy crystal/solution. Part IV explains how the kinetic effect of confinement due to micrometer size, and so nanovolumes, leads to isolation of different phases, that are characterized by X-Ray Diffraction (XRD).

## I. Properties of microfluidic devices

Microfluidic crystallization of biomolecules takes place in micrometer-sized channels or wells (tens to few hundreds of micrometers). Hence, crystals form in droplets or wells/reservoirs, that are of nanometric volumes. The physical properties of microfluidic devices are linked to hydrodynamic and mass transport parameters at micro-nanometric level (Zhang et al. 2008, Kumar et al. 2011, Osouli-Bostanabad et al. 2022). In the case of biomolecules, materials used to fabricate the devices are of great importance for storage of crystals and for their in-situ analysis (see chapter 2 of part A of the present book).

### I.1. Materials for the fabrication of microfluidic devices

Many materials can be used in the fabrication of microfluidic systems. Materials used for fabricating microfluidic devices can be divided into 2 categories:

(i) Inorganic materials, as glass;

(ii) Thermoplastics and elastomers as SU-8, polydimethylsiloxane (PDMS), Cyclic Olefin Copolymer (COC), Polymethylmethacrylate (PMMA), PolyEther Eher Ketone (PEEK), poly Fluoropolymers such as Perfluoroalkoxy (PFA), Fluorinated Ethylene Propylene (FEP), polytetrafluoroethylene (PTFE), Tetrafluoroethylene, hexafluoropropylene and vinylidene fluoride (THV).

Table 1 shows the materials used according to the method of crystallization. The material mainly used is PDMS due to its ease for sealing without distorting channels geometry (Shi et al. 2017). It can be added a glass layer to a PDMS chip, using a sol-gel process, in order to increase resistance to organic solvents (Abate et al. 2008). Furthermore, composite materials such as COC–PDMS presents promising properties for microfluidic applications due to their unique physical, chemical, and optical properties.

**Table 1** Materials used in microfluidics according to the method of crystallization: associated devices, biomolecules with objectives and techniques of analysis

| Methods | Specificity | Devices name | Material | Property | Applications with biomolecules | Objectives | Analysis | Ref |
|---|---|---|---|---|---|---|---|---|
| Vapor-diffusion derived methods | Passive solid layer | Osmosis-based | PDMS | Permeability to water | Monoclonal antibody anti-CD20 | Increase the mean size of crystals by crystallization/dissolution cycles | Ex-situ X-Ray | Morais et al. 2021 |
| | Passive solid layer | Phase chip | PDMS | Permeability to water | Xylanase | Screening crystallization conditions | Ex-situ X-Ray | Shim et al. 2007; Selimovic et al. 2009 |
| | Active solid layer | Micro-Contactor | PDMS/ion-exchange membrane | Permeability to water and ions | Lysozyme | Decoupling nucleation and growth of lysozyme | Ex-situ X-Ray | Polino et al. 2021 |
| | Thin liquid layer | Osmotic-based | PDMS/Glass composite | Diffusion of water | Lysozyme | Crystals with a high diffraction quality by derivatization | In-situ X-Ray | Zheng et al. 2004 |
| Dialysis derived methods | Semi-permeable membrane | NOA81 based micro-dialysis | Cellulose membrane / NOA 81 /PMMA | Permeability to precipitant | Lysozyme, IspE, and insulin | Good transparency to X-rays for in situ measurements | In-situ X-Ray | Junius et al. 2020 |
| Counter-Diffusion derived methods | FID Devices with valves | Barrier Interface Metering (BIM) | PDMS | Pneumatic external valves | Lysozyme, bacterial primase catalytic core domain, type II topoisomerase ATPase domain/ADP, Thaumatin, Xylanase and Glucose isomerase | Large single crystals | Ex-situ X-Ray | Hansen and Quake 2003 |
| | | Centrifugally-actuated device | SU-8 polymer | Centrifugal valves | Lysozyme as a model system and calcium–calmodulin dependent kinase II (CaMKIIβ) | High level transparency to X-rays | In-situ X-Ray | Saha et al. 2023 |
| | FID Devices without valves | SlipChip | Glass | Lubricating layer of fluorocarbon | Photosynthetic reaction center from Blastochloris Viridis | Screening crystallization conditions | Ex-situ X-Ray | Du et al. 2009 |
| | | ChipX3 | PDMS or PMMA or COC | Tree-network | Enzyme, protease, nanobody, lipase, aspartyl-tRNA synthetase, Mitochondrial aspartyl-tRNA synthetase, OMT Shua, RNA duplex and hemoglobin | High level transparency to X-rays | In-situ X-Ray | De Wijn et al. 2019 |
| Batch derived droplet-based methods | With surfactant | Co-Flow Emulsion | PDMS | Rectangular capillary | Lysozyme | Nucleation rate | Ex-situ X-Ray | Akella et al. 2014 |
| | | Two-dimensional screening | PDMS | Thermal gradient | Adipic acid | Solubility curve | Ex-situ X-Ray | Laval et al. 2007 |
| | | Hybrid method | PDMS chip and Teflon (PFA, PFE, PTFE) capillary | Droplet-size varying with precipitant concentration | Model membrane proteins: Reaction center (RC) from *R. viridis* and Porin from *R. capsulatus* | Simultaneous screening and optimization | In-situ X-Ray | Li et al. 2006 |
| | Without surfactant | Easy-to-use Chip | PDMS and Teflon capillary | Hundreds droplets for statistical approach | Lysozyme Size and frequency of droplets | Fundamentals of nucleation: solubility curve, phase diagram, nucleation rate, interfacial energy | Ex-situ X-Ray | Zhang et al. 2015 |
| | | Plug and play device | Teflon (PFA, PFE, PTFE) capillary and PEEK junction | | in aqueous and viscous solutions: lysozyme, rasburicase, human quinone reductase 2 (QR2), sulfathiazole and paracetamol in ethanol: caffeine and isonicotinamide in nitrobenzene: isonicotinamide in isopropanol: glyclazide in acetonitrile and isopropanol: paracetamol | Screening an optimization Polymorphism In-situ Raman and X-ray measurements Fundamentals of nucleation: solubility curve, phase diagram, nucleation rate, interfacial energy | Ex-situ X-Ray In-situ X-Ray Raman | Lambert et al. 2023 |

The choice of material is an important step, which is notably determined by cost, fabrication delay, equipment and technologies available, and above all the nature of the phases used. Depending on the crystallization solvent (aqueous or organic) and experimental conditions (duration, size of channel, range of temperature, range of pressure), the choice of material is guided by the following physico-chemical properties:
- optical transparency for direct observation via microscopy
- compatibility to aqueous or/and organic solvents
- wetting properties in relation to water - hydrophobicity allows the formation of aqueous solution droplets in an oil as continuous phase, and hydrophilicity the opposite.
- surface modification to enhance compatibility to solvents or change wetting properties
- roughness which affects fluid velocity along the channel resulting in larger flow resistance, and so decrease of fluid velocity
- thickness due to the use of valves to close/open chambers or membranes to let evaporation of solution or exchange of ions/precipitants
- transparency to UV, Raman and X-ray lights, to control concentration and crystal structure.

Each of these materials has its advantages and its own limitations. Hence by combining two materials, microfluidic devices can be improved for application in crystallization studies.

**I.2. Transparency of materials for in-situ analysis of crystals**

*I.2.1 Materials for Raman analysis*

Fiber-optic Raman systems were used in microfluidics with a monochromatic beam (laser) directed onto the sample at a wavelength of 785 nm (Lambert et al. 2023). The scattered signal is then collected and filtered to recover only the wavelengths corresponding to the Raman signal. For instance, the group of Berglund used Raman spectroscopy to monitor lysozyme concentration during crystallization in a hanging drop experiment in real time (Schwartz and Berglund 1999, Tamagawa and al. 2002).

For in-situ analysis, the material of the microfluidic device is crucial. Therefore, our group measured Raman spectra on capillaries made of different types of glass (conventional, borosilicate and quartz glass) and Teflon PFA (0.5 or 1 mm inner diameter and 1.6 mm outer diameter) (Fig. 1). We used a QEPro Raman spectrometer (Ocean Insight) under laser excitation at 785 nm. The Raman spectra of the different types of glass (classical, borosilicate and quartz glass) do not show defined peaks, but halos, indicating that these materials fluoresce. This makes these materials unsuitable for on-chip analysis. In contrast, Teflon PFA shows well-defined peaks, more intense in the case of the thicker-walled 0.5 mm ID capillary. As a result, Teflon PFA capillaries can be used to analyze droplets using Raman spectroscopy, provided that the molecules analyzed show peaks at other Raman shift values than those of PFA (Lambert, 2023).

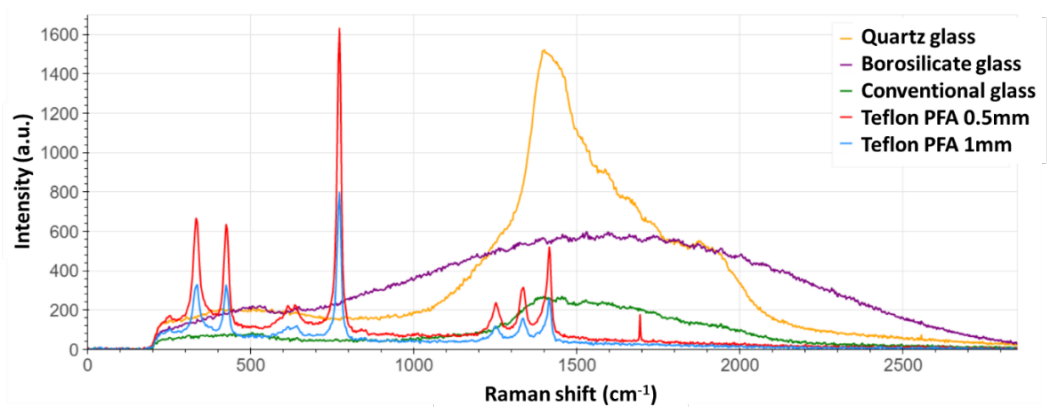

**Fig. 1** Raman spectra of quartz glass, borosilicate glass, conventional glass and Teflon PFA capillaries with internal diameters of 0.5 mm and 1 mm measured at 785 nm with the QEPro Raman spectrometer after (Lambert, 2023)

*I.2.2. Materials for X-ray analysis*

The interactions of X-rays with matter are mainly absorption and scattering, as described in chapter 2. Due to the material of device that is in the path of the X-ray beam, absorption induces attenuation of the radiation that reaches the crystal in solution (Sui et al, 2016). The nature of the materials and its thickness are highly relevant parameters since attenuation can reduce the intensity of both the incident and diffracted X-rays (Sui et al, 2016, Sui and Perry 2017, Guha aet al. 2012). Scattering effects from the materials of the device add up to X-ray scattering induced by the presence of air surrounding the experiment and/or the presence of fluids within the crystal structure and lead to increased background noise (Sui et al, 2016, Sui and Perry 2017, Guha aet al. 2012). Therefore, decreasing the total thickness of the chip or using low density materials can moderate the impact of attenuation and scattering on the diffraction measurements. Nonetheless, decreasing the thickness of such devices also gives rise to some limitations concerning their mechanical stability and the fabrication process could be complicated (Sui and Perry 2017). Table 2 shows X-ray interactions with materials through background images (Dhouib et al. 2009) or background scattering (Junius et al 2020).

Table 2 Comparison of material X-ray scattering properties: a) Background images produced in the absence (i.e. in air only) and in the presence of materials. Reprinted with permission from Ref. (Dhouib et al. 2009). Copyright 2009 Royal Society of Chemistry; b) Background scattering intensity as a function of resolution (Å) from the materials composing a microfluidic device. Reprinted with permission from Ref. (Junius et al 2020). Copyright 2020 Royal Society of Chemistry

| Groups | Materials | Thickness | Content | X-Ray scattering |
|---|---|---|---|---|
| **Background images: at a wavelength of 1.54 Å from Sauter group [17]** | Air | - | H, C, O, N | 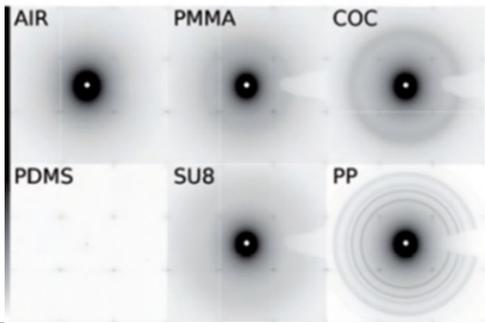 a) |
| | Polyméthacrylate de méthyle (PMMA) | 230-250 µm | H, C | |
| | SU8 photoresist | | | |
| | Cyclic olefine copolymère (COC) | | H, C, O | |
| | Polypropylene (PP) | | | |
| | Polydiméthylsiloxane (PDMS) | 1 mm | H, C, O, Si | |
| **Background scattering: Intensity versus resolution Budayova-Spano group [18]** | Kapton tape (polyimide) | 20 µm | H, C, O, N | 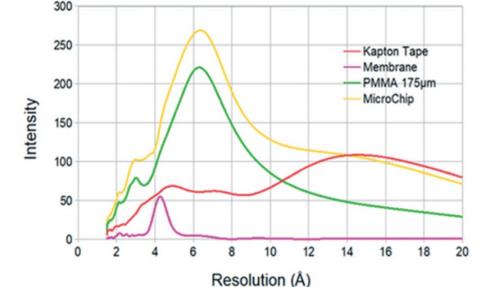 b) |
| | Regenerated cellulose (RC) dialysis membrane | 40 µm | | |
| | Polyméthacrylate de méthyle (PMMA) | 350 µm | H, C | |

For background images, the thickness of PDMS was limited to a minimum value, below which it is too flexible for handling. All polymers containing only light atoms (C, O, N, H) exhibited scattering backgrounds that were comparable in intensity to air (Table 2a). PDMS contains silicium, rendering it not suitable for the analysis by XRD. Furthermore, PP was the only material which showed discrete ring-like patterns characteristic of a micro-crystalline structure.[17] For the study of an entire device, background scattering of each part of the device shows its proper intensity versus resolution (Table 2b). The large and wide peak of PMMA corresponds to the diffusion of X-rays in the material. The peak is more diffuse for Kapton and even more for the RC dialysis membrane tape, apart a peak at 4-5 Å. However, the noise generated from the entire microchip is maximum (peak) at low resolution (6–7 Å), which is less critical for processing the data at higher resolution. Moreover, this background noise is 15 to 27 times lower than the one of commercial crystallization plates (Greiner CrystalQuickX and Crystal Cube by Cube Biotech) used for diffraction experiments at room temperature (Junius et al 2020).

## II. Microfluidic devices and methods of crystallization

In bulk crystallization, methods of crystallization used to increase the concentrations of biomolecule and/or precipitant in solution drives crystals formation and their quality. To simplify, Chayen et al. drew a typical phase diagram for protein crystallization, only considering the variation of protein and precipitant concentrations (Fig. 2) (Chayen and Saridakis 2008):

- Batch methods, where a solution with known quantities of protein and precipitant solution corresponds to a single condition (Fig.2.1);
- Vapour-diffusion methods, where a mixture of protein and precipitant is placed in vapour contact with a reservoir of precipitant solution and allowed to equilibrate (Fig. 2.2);
- Dialysis methods, where protein and precipitant solutions are separated by a semipermeable membrane that allows the precipitant to enter the sample while maintaining the concentration of the protein (Fig. 2.3);
- Counter-diffusion methods, where the protein and precipitant diffuse into each other, sampling a wide range of concentration space (Fig. 2.4)

For all of the methods displayed in Fig. 2, crystal growth after nucleation consumes molecules of the solution decreasing the concentrations of protein, according to the method.

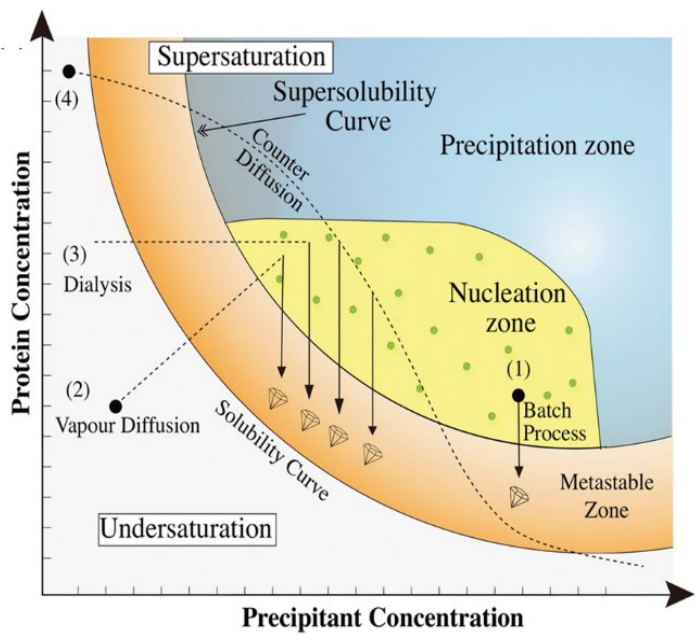

**Fig. 2** (a) A schematic phase diagram showing typical crystallization methods: (1) batch method; (2) vapour-diffusion method; (3) dialysis method; (4) counter-diffusion. Reprinted with permission from Ref. (Saha et al. 2023) redrawn from Chayen et al. (2008) Copyright 2023 Royal Society of Chemistry

In microfluidics, crystallization experiments follow the same kind of path in the phase diagram as described in Fig. 2, until reaching the nucleation zone far above the solubility curve. For vapor-diffusion, dialysis and counter diffusion methods, different conditions of biomolecule and/or precipitant concentrations are screened in a single experiment. Whereas in the case of batch methods the solution is directly carried in the nucleation zone generally by changing the temperature. Hence, screening different crystallization conditions requires several experiments. Therefore, these two categories of methods are distinguished in this review and some representative examples of devices used for each crystallization method of Fig. 2 (Table 1) are presented (non exhaustive). These devices can result in very complex ones for high-throughput in-situ crystallography, that is treated in more details in chapter 2 of part A of the present book.

## II.1. Vapor-diffusion derived methods

The simplest method used to screen crystallization conditions is the hanging (or sitting) drop vapor diffusion method. This technique consists of depositing a drop containing a mixture of biomolecule and precipitants (salt, polymer, pH reagents) close to a reservoir containing a higher concentration of precipitants. The difference in osmotic pressures of the two solutions is due to the difference in their chemical activities. Hence, they are balanced by a slow diffusion of water from the crystallization solution into the reservoir inducing a decrease in volume of the drop and so concentrating its components (biomolecule and precipitant). Therefore, a single crystallization trial proceeds through a range of conditions in components concentrations, thereby conducting a self-screening process, until the appearance of crystals (Fig. 2).

In microfluidics, vapor-diffusion derived methods use a permeable layer separating a small volume of crystallization solution, with biomolecule and precipitant at a concentration typically undersaturated, and a solution highly concentrated in precipitant but without biomolecule. The small volume of crystallization solution equilibrates with the solution of precipitant, causing increase of biomolecule and precipitant concentrations in a large range. This method is also called Osmosis. The layer separating the crystallization solution (with biomolecule) from the reservoir of precipitant solution (without biomolecule) can be solid or liquid and it can be active or passive.

### II.1.1. Passive solid layer

- Salmon group designed a Osmosis-Based Microfluidic Chip, which is a simple two-level chip with a thin PDMS layer separating plugs (linear droplets) from an open microfluidic reservoir. Plugs containing a mixture of protein and precipitant were stored in a chip and the reservoir imposed a given water chemical activity (Fig. 3) (Morais et al. 2021). Because PDMS is only permeable to water, the imposed water activity therefore increases or decreases in a controlled way the concentration of protein in the droplet and consequently the supersaturation of the solution. Moreover, the small length of the channel containing the plug (h = 30 μm) ensure perfectly controlled mass transport conditions preventing for instance unwanted free convection for crystal growth that can occur in hanging drops (Shim et al. 2007a). Finally, the specificity of their approach is that the concentration of all the solutes is known at any time from the length of the plug, thus allowing the extraction of quantitative information on the crystallization process.

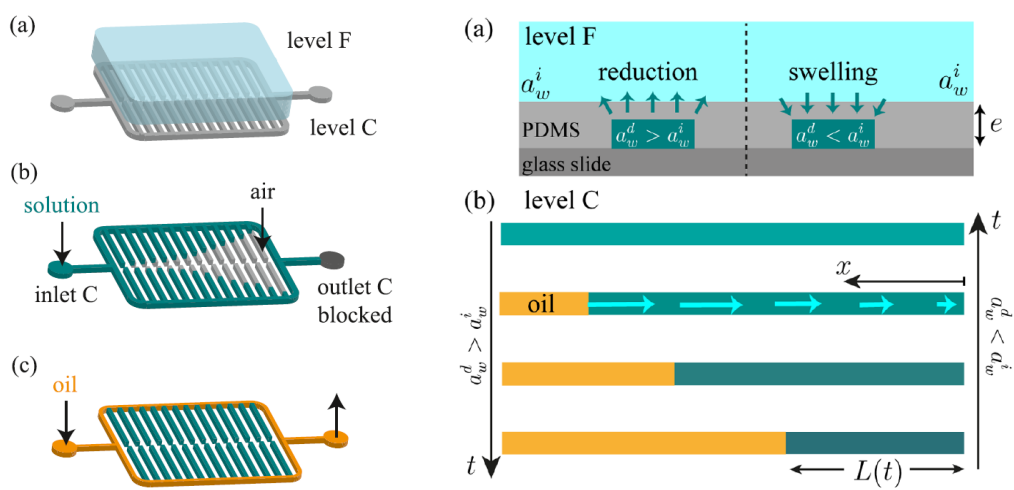

**Fig. 3** Osmosis-Based Microfluidic Chip : on the left, protocol of (a) filling of level F with a solution imposing a given water chemical activity, (b) filling of level C with plugs of the (protein + precipitant) solution (outlet C is closed), and (c) flow of oil from inlet C to outlet C to store the plugs in the channels without contact with air; on the right, (a) cross-sectional view and arrows showing the water flux to balance the water chemical activities between levels C and F, increasing the concentrations in the drop or diluting the drop, (b) top view evidencing decrease or increase of the plug length. The arrows show the permeation-induced flow v(x) in the plug with time in the case of volume reduction. Reprinted with permission from Ref. (Morais et al. 2021). Copyright 2021 American Chemical Society

The authors exploit this chip on a protein of therapeutic interest the full-length monoclonal antibody anti-CD20. They show that the fine-tuning of the permeation rate makes it possible to perform crystallization/dissolution cycles to selectively dissolve small crystals and increase the mean size of the remaining crystals via a kinetic ripening mechanism. Such a kinetic control of the crystallization process opens the possibility to obtain crystals with a sufficient size to determine their structure by X-ray diffraction measurements.

- Thin layers in PDMS were also used for very complex systems to screen crystallization conditions: Fraden group developed a device denoted the *Phase Chip* to measure and manipulate the phase diagram of multi-component fluid mixtures (Fig. 4) (Shim et al. 2007a, Selimovic et al. 2009). The device is constructed from two PDMS layers and subsequently sealed together (Fig. 4b). In the upper, thick (5 mm) layer there are flow channels and storage wells. In the lower, thin (40 μm) layer there is a reservoir sealed by a 15 μm thick PDMS membrane. The reservoir is formed by spin coating a 40 μm thick layer PDMS over a 25 μm high photoresist mold (Squires and Quake 2005).

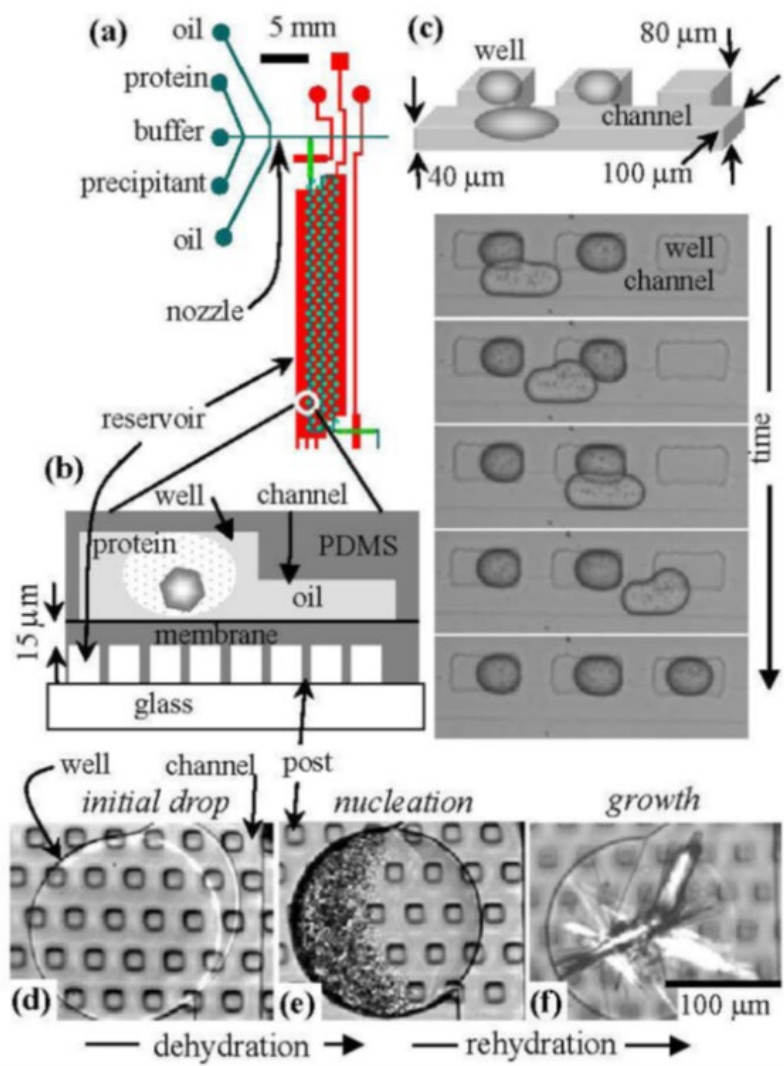

**Fig. 4** (a) Plan view of one of the five sections of the Phase with a reservoir (red) located underneath 100 wells (green circles); (b) Vertical section of a storage well, channel, reservoir and dialysis membrane. (c) Storage of droplets into rectangular wells. (d-f) Protein crystallization with reversible dialysis. The photographs are of a single 300 μm diameter circular well that contains protein solution. The channel is on the right side of the well. The square posts are 30 μm wide and support a 15 μm thick PDMS membrane, which forms the bottom of the well. (d) A stable protein solution of xylanase slightly overfills the well. (e) Protein gelation occurred after the reservoir was filled with salt solution. (f) The reservoir was filled with pure water, which rehydrated the precipitate, transforming them into crystals. Reprinted with permission from Ref. (Shim et al. 2007a)

A droplet that partially occupies both a channel and well will experience a gradient in surface energy, with the resulting force acting to drive and store the droplet inside the well. As the wells exist as pockets on the sides of the channel, the enclosed, stored droplets are outside the flow stream and shielded from dislodgment by hydrodynamic forces. Droplets sequentially fill the wells, with the first drop going into the first well. Subsequent drops pass over all filled wells, entering the first empty well. To prevent coalescence of the drops during the loading process, surfactants must be added to the continuous phase.

For crystallization of xylanase, droplets of the dispersed phase (D) containing xylanase and precipitant at different compositions were generated in a continuous phase (C) called oil in Fig. 4a. Then, these droplets circulated in the channel until they fill the wells (Fig. 4c). Finally, the permeation of water through PDMS was used to controllably vary the concentration of xylanase in wells (Fig. 4b) in order to obtain crystals (Shim et al. 2007a).

The authors also employed the Phase Chip for statistical studies of lysozyme crystal nucleation, by decoupling nucleation and growth of crystals and so they improve their yield and quality. Therefore, they reversibly varied the supersaturation of lysozyme inside the stored droplets by controlling the chemical potential of the reservoir (Selimovic et al. 2009).

### II.1.2. Active solid layer

Polino group used an ion-exchange membrane (117 Nafion®) in their microfluidic contactor to separate two independent chambers in PDMS: a wells layer dedicated to the crystallization solution and a channels layer filled with stripping/derivatization solution (Fig. 5) (Polino et al. 2021). The membrane provides controlled transport of water and ions, and is used in this work to facilitate the derivatization of protein crystals with heavy atoms, i.e. the incorporation of heavy atoms into the crystal to maintain the isomorphism for obtaining phases for unknown structures and for low-resolution data sets.

This device proved to allow for a successful crystallization of lysozyme and subsequent in situ and gentle crystal derivatization based on an efficient control of the diffusion rates of the derivatizing agent ($Hg^2$). Thus, the crystal structural integrity was assured along the derivatization process, proving to be a more efficient alternative to the traditional crystal soaking methodologies.

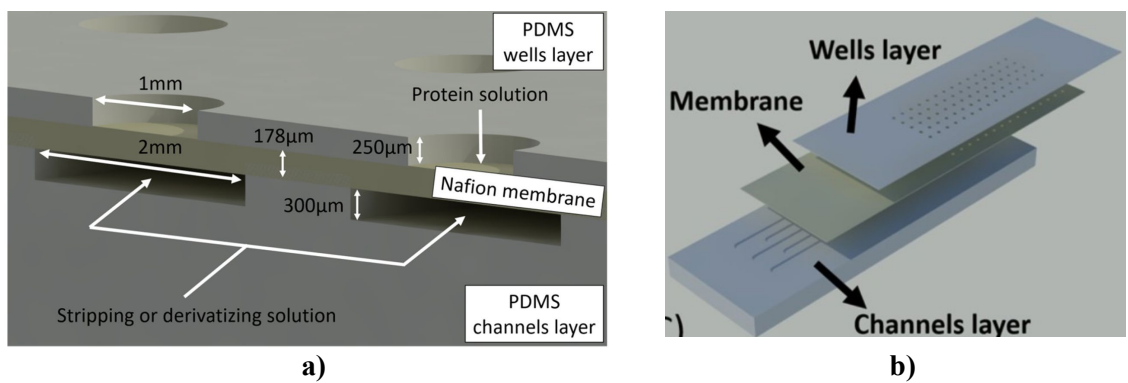

**Fig. 5** Microfluidic Contactor with Nafion®117 Membranes: a) Cross-section scheme of the microdevice; b) AutoCAD image showing the different components of the microdevice, i.e., channels layer (on the bottom), membrane (in between layers), and the wells layer (at the top). Reprinted with permission from Ref (Polino et al. 2021). Copyright (2021) MDPI

### II.1.3. Passive liquid layer

Ismagilov group developed a composite PDMS/Glass capillary microfluidic system in which they generated droplets of dispersed phases in a permeable oil (polytrifluoropropylmethylsiloxane homopolymer-FMS-121) as continuous phase (Fig. 6a) (Zheng et al. 2004). This permeable oil played the role of a thin liquid layer enabling diffusion of water from a droplet to another. Therefore, they generated alternating droplets: sequence of droplet 1 containing protein and precipitants and droplet 2 containing a high-concentration of salt. To avoid evaporation in PDMS during crystallization trials, they push the droplets in a glass capillary to the PDMS channel. The flow of oil transported droplets into a glass capillary attached to the outlet of the PDMS section. The flow was stopped when the glass capillary was filled with droplets of desired composition, and it was disconnected, sealed with wax, incubated at 18 °C, and monitored periodically. In contrast to droplets incubated in the PDMS channels, droplets and crystals incubated in a sealed glass capillary were stable even in the absence of humidity control, and did not show signs of evaporation over six months. During storage, water diffuses from the droplet 1 towards the neighboring droplet 2 until osmotic pressures were equalized. Then, solute and precipitant concentrations in droplet 1 increased until reaching a supersaturation high enough to start nucleation of crystals. The authors screened and optimized lysozyme crystallization conditions (Fig. 6b). Moreover, the thin-walled glass capillary permitted to obtain diffraction patterns from crystals by subjecting the capillaries with crystals directly to the synchrotron X-ray beam.

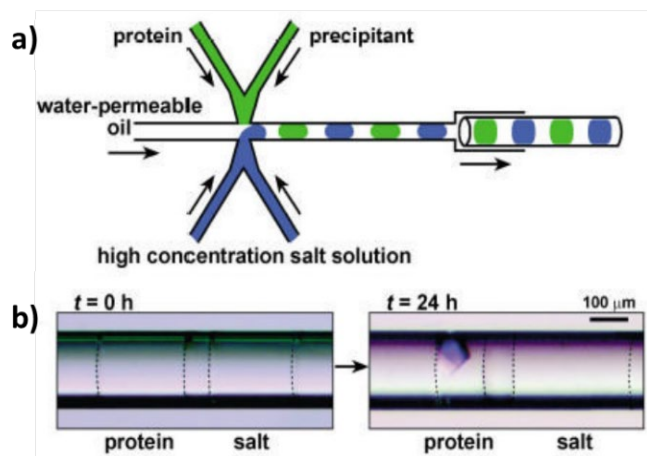

**Fig. 6** a) A schematic illustration of the PDMS/glass composite device for setting up pairs of droplets for protein crystallization under vapor-diffusion conditions in capillaries. b) Two microphotographs that show a pair of droplets immediately after (left) and 24 h after (right) the setup. The droplet of the protein solution yielded a crystal after it had lost about 50% of its water content. Dashed lines were added to show the interface between the aqueous droplets and water permeable oil. Reprinted with permission from Ref. [Zheng et al. 2004]. Copyright 2004 John/Wiley & Sons

## II.2. Dialysis derived methods

In dialysis method, the concentration of biomolecule is maintained constant while the concentration of precipitant is increased. Therefore, solution of biomolecule and solution of precipitant are separated by a semi-permeable membrane that is impermeable to biomolecule but that allows passage of precipitant. The crystallization trial, on the side of the membrane containing the protein, starts at a low precipitant concentration. The precipitant diffuses across the dialysis membrane into the biomolecule solution, causing an increase in precipitant concentration (Fig. 2).

For micro-dialysis using microfluidics, Budayova-Spano group used semipermeable membranes with a molecular weight cut-off smaller than the dimensions of the biomolecules but larger than the precipitant, polyethylene glycol (PEG). They developed a microfabrication process enabling the integration of regenerated cellulose dialysis membrane between two layers of a microchip in NOA 81 resin on a PMMA substrate (Fig. 7) (Junius et al 2020).

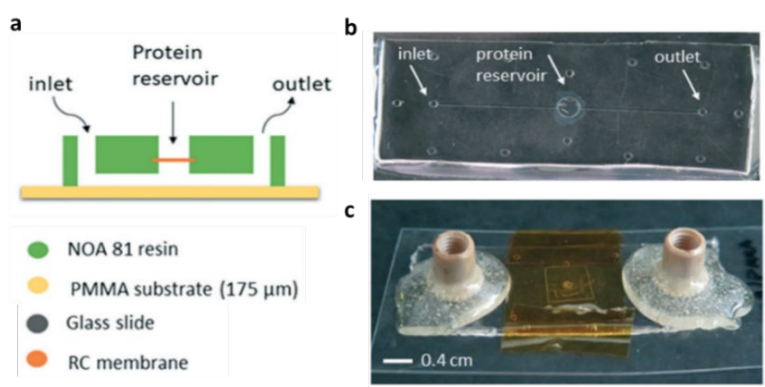

**Fig. 7** NOA81 microchip: a) schematic illustration of the microfluidic chip embedding a regenerated cellulose dialysis membrane (transverse view); b) top view of the chip with the inlet/outlet for the fluidic connectors and the protein reservoir (top view); c) encapsulation of the biomolecule solution in the central reservoir by a piece of PMMA and Kapton tape, Peek connectors bonded on the inlet and outlet of the device with fast epoxy glue. Reprinted with permission from Ref. (Junius et al 2020). Copyright 2020 Royal Society of Chemistry

The protein solution is encapsulated within the dialysis reservoir, leading to a crystallization experiment under static conditions regarding the biomolecule solution. However, the solution of PEG flows continuously within the microfluidic channel, offering the possibility to exchange dynamically crystallization conditions. PEG diffused across the membrane due to the difference in concentration between the two compartments separated by the dialysis membrane.

Micro-dialysis method is reversible as is the action of temperature. Thus, knowing the phase diagram, the authors used their microfluidic chip for optimizing crystallization conditions by mapping temperature–precipitant concentration phase diagrams for lysozyme. Hence, high quality large crystals of three proteins were obtained: lysozyme, IspE, and insulin for in situ X-ray diffraction measurements. The microchip assembly ensured the absence of leakages from the reservoir and shew good transparency to X-rays. Therefore, the microchip was mounted onto the beamline and partial diffraction data sets were collected in situ from several isomorphous crystals and were merged to a complete data set for structure determination (Junius et al 2020).

## II.3. Counter-Diffusion derived methods

Counter-diffusion method involves the placement of biomolecule solution adjacent to that of the precipitant whereby the solutions are initially in contact to one another. In a restricted geometry, like in microfluidics, crystallization process is allowed to proceed by diffusively mixing the protein and precipitant solutions (Otalora et al. 2009). Counter-diffusion is used in microfluidics because it combines the benefits of slow diffusive mixing and the possibility of screening crystallization conditions. Indeed, gentle and well controlled diffusive mixing of protein and precipitant solutions, facilitating the crystallization process, is observed (Dhouib et al 2009, Emamzadah et al. 2009, Ng et al. 2008, Ng et al. 2003). Due to the difference of molecular weight between the protein and the precipitant, this latter diffuses more quickly than the protein. As the diffusion front moves through the capillary, the concentration gradient sample a range of supersaturation (ratio between the concentration and the solubility). Hence, crystals are numerous and small near the contact zone (high supersaturation), and few and larger far from the contact zone (low supersaturation). Therefore, counter-diffusion provides screening capabilities for a wide range of supersaturation in a single experiment (Fig. 2).

There are several implementations that enable two solutions to mix by diffusion, usually called free-interface diffusion (FID) or liquid-liquid diffusion. The solution of biomolecule and the solution of precipitant can be separated by an isolation valve, by centrifugation or in branch lines.

### II.3.1. FID Devices with valves

Pressure-driven pneumatic valves are often used in microfluidic devices and when the valve is opened, the solution of biomolecule and the solution of precipitant counter-diffuse, as indicated in Fig. 8 for proteins (Saha et al. 2023).

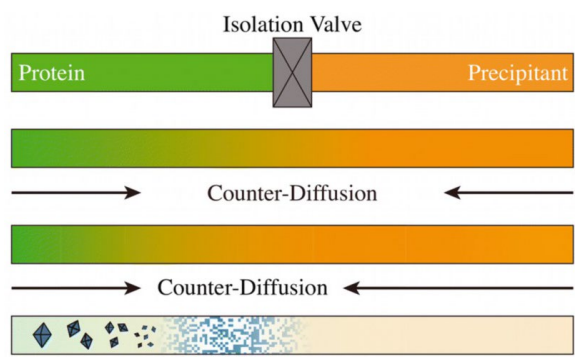

**Fig. 8** Schematic depiction of a counter-diffusion experiment. An isolation valve separates protein (green, left) and precipitant (orange, right) solutions in a capillary. Upon opening of the valve, the two solutions counter-diffuse, as indicated by the color gradient. Near the contact line between the two solutions, the concentration of precipitant is so high that it increases supersaturation of protein leading to precipitation. Further forward from the contact zone, the supersaturation of protein due to the presence of precipitant is lower leading to less crystals with larger size. Reprinted with permission from Ref. (Saha et al. 2023). Copyright 2023 Royal Society of Chemistry

- To integrate valves and pumps, Quake group used micromechanical systems fabricated in silicone elastomer by soft lithography process, that they incorporated into their device. They constructed a unit cell designed to conduct three FID experiments at different fluid mixing ratios that they called Barrier Interface Metering (BIM) (Fig. 9) (Hansen and Quake 2003). Their BIM contained three pairs of microfluidic chambers with well-defined volumes. Each pair of chambers displayed two containment valves situated at chambers entrances for the filling with solutions and one barrier valve for the control of the fluidic interface between the chambers. The authors implemented 48-unit cells in a single crystallization chip, enabling 144 simultaneous FID protein crystallization reactions (Fig. 9d).

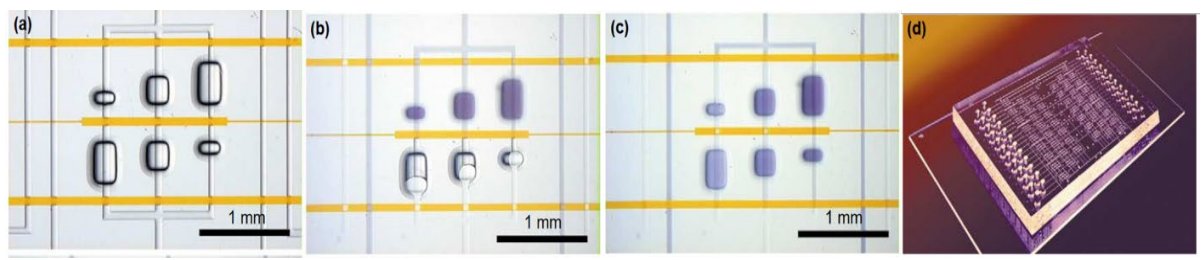

**Fig. 9** Barrier Interface Metering (BIM) with three pairs of microfluidic chambers with well-defined volumes: a) microfluidic unit cell designed to implement three free interface diffusion assays at three mixing ratios; b) First, the barrier valves are closed and the containment valves are opened, one chamber is primed with protein solution and the other one with precipitant solution using pressurized outgas priming; c) Containment valves are closed isolating the unit cell and barrier valve opened allowing diffusive mixing between the coupled chambers; d) a microfluidic chip with 48 of the unit cell structures (BIM) shown in a) performing 144 simultaneous FID crystallization experiments using only 3 mL of protein sample. Reprinted with permission from Ref. (Hansen and Quake 2003). Copyright 2003 Elsevier

Because the amount of each component participating in the reaction is precisely defined by the volumes of the microwells, this technique allows control over the end point of each reaction. Moreover, by shortening the barrier channels through which diffusion of protein and precipitants occurred they could reduce the number of crystals per well. In the plethora of studies realized by the authors, they studied crystallization of many biomolecules such as lysozyme, bacterial primase catalytic core domain, type II topoisomerase ATPase domain/ADP, Thaumatin, Xylanase and Glucose isomerase (Hansen et al. 2002). As they could grow large single crystals, these could be harvested directly from chip reactions and used for high-resolution diffraction studies (Hansen and Quake 2003).

- To avoid the use of external pumps, Perry group developed a centrifugally-actuated microfluidic device with centrifugal valves (Saha et al 2023). They created a multi-layered device by combining photolithography and nanoimprint lithography (Fig. 10). They used SU-8, a UV curable polymer, for its high transmission of X-Ray in contrast to the silicone-based elastomers more typically used in microfluidic devices that would attenuate nearly 53 % of the signal for the same thickness of material. The device consisted of 8 identical sections, as shown in Fig. 10a, which could each run independent experiments and be easily separated from the others by cutting along a series of perforated lines and mounting onto a magnetic base for X-ray analysis. A single device included three openings (Fig. 10b) that were connected to the crystallization chamber by a long channel containing a U-turn. The centrifugal force acted on the solutions on both sides of the U-turn to split them and make them flow separately into the inlet and the crystallization chamber (Fig. 10 c2 and c3). This facilitated the reproducible metering of a known volume of liquid (160 nL) into the crystallization chamber.

Liquid handling took advantage of surface forces to control fluid flow and enable metering, without the need for pressure pump and connecting tubes. In addition, wetting properties is important in this device: the hydrophilic nature of the SU-8 polymer facilitated easy filling of the device by capillary action; the deposition of layer of fluorinated alkyl silane on the bottom substrate of the centrifugal valve rendered it hydrophobic to stop the flow of liquid before centrifugation (Fig. 10e).

When centrifugal force was applied, the two solutions enter the crystallization chamber to create a counter-diffusion setup. The authors showed that a spin speed of 1500 rpm allowed for gentle introduction of liquids into the crystallization chamber yielding counter-diffusive mixing. Thus, they used their device for crystallization and in situ, room temperature, structural analysis of lysozyme as a model system and calcium–calmodulin dependent kinase II (CaMKIIβ) to determine its hub domain. Because they used thin UV-curable polymers that are highly transparent to X-rays, they were able to perform X-ray crystallography in situ, eliminating the manual handling of fragile protein crystals and streamlining the process of protein structure analysis.

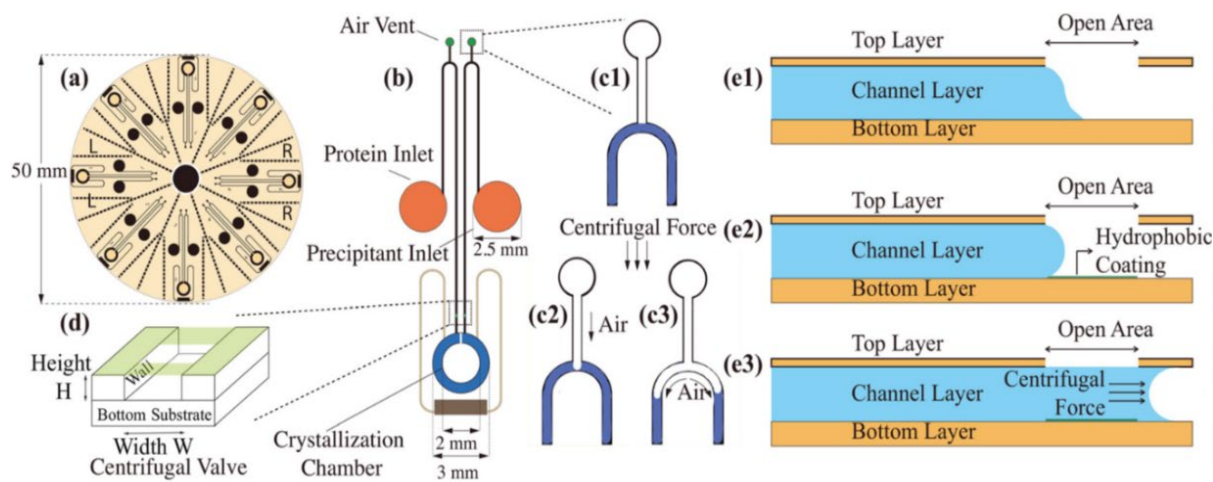

**Fig. 10** Centrifugally-actuated microfluidic device: (a) Schematic of the overall device design containing 8 identical sections. (b) Schematic of each individual section depicting 3 inlet holes for the introduction of solutions, air vents and centrifugal solution respectively (c) Schematic depicting the function of the air vent located at the top of the U-turn facilitating metering of solutions (c1). Action of the centrifugal force on both sides of the U-turn and addition of air vent enabling for the creation of a meniscus at the peak of the U-turn (c2 and c3). (d) 3D view of the centrifugal valve located right before the crystallization chamber to stop the flow of liquid and (e) a 2D cross-sectional view showing the role of surface forces and wetting related to the operation of the centrifugal valve. Reprinted with permission from Ref. (Saha et al 2023). Copyright 2023 Royal Society of Chemistry

In conclusion, the use of valves in FID requires external control equipment, and valves are often composed of PDMS. PDMS devices have the additional complication of requiring control of the atmosphere and evaporation. In the case of device in SU8 with centrifugal valve, the presence of air in the crystallization chamber and the treatment of valve surface can be drawbacks.

*II.3.2. FID Devices without valves*

- Ismagilov group developed a microfluidic device designed to perform multiplexed microfluidic reactions without pumps or valves, called SlipChip (Du et al. 2009). This device was composed of two plates in close contact fabricated in glass (Fig. 11). The sample-containing wells of the top plate was exposed to the reagent-containing wells of the bottom plate (Fig. 11e), and facilitated diffusion and reactions. Slipping of the two plates was facilitated by using a lubricating layer of fluorocarbon. Using the SlipChip, the authors screened the crystallization conditions of a membrane protein, the photosynthetic reaction center from *Blastochloris viridis,* against two sets of 24 precipitants in duplicate (disposed in 4 rows). They observed that crystallization in the same precipitant (($NH_4$)$_2SO_4$) but with increasing concentrations leads to precipitation.

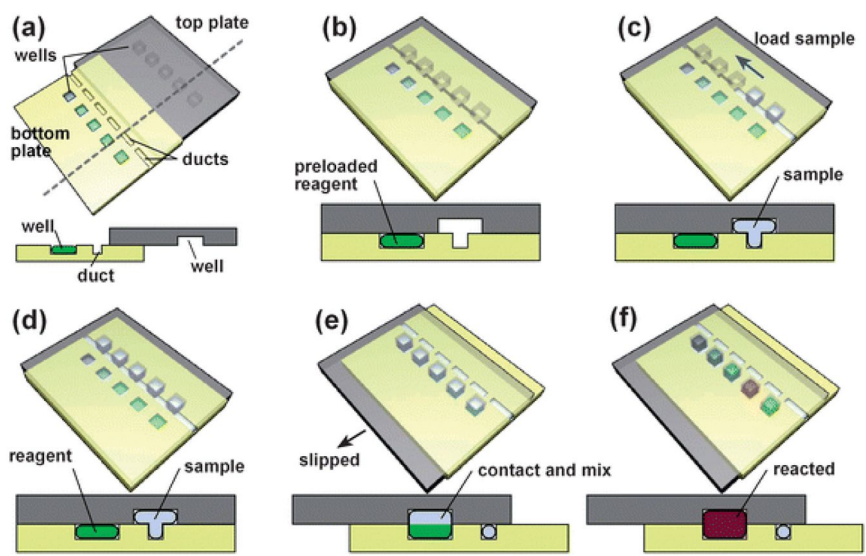

**Fig. 11** 3D schematic drawings with cross-sectional views that describe the operation of the SlipChip: a) o The bottom plate contained an array of wells preloaded with many reagents and an array of disconnected ducts for subsequent loading of the sample (the solution of biomolecule); b) The top plate served as a lid for the wells of the bottom plate and also contained an array of wells that were complementary in pattern to the array of ducts in the bottom (c) and (d) the ducts of the bottom plate are connected into a continuous fluidic path, in which the sample is loaded; e) the top plate was slipped relative to the bottom plate so the complementary patterns of wells in both plates overlapped, exposing the sample to the reagents; f) The red well schematically shows a reaction taking place after mixing and incubation. Reprinted with permission from Ref. (Du et al. 2009). Copyright 2009 Royal Society of Chemistry

- With the aim of using a simpler counter-diffusion chip, Sauter group designed a chip geometry composed of eight crystallization channels which were connected through a dichotomic tree-like network on one side to a single inlet or well (Fig. 12) (de Wijn et al. 2019). The geometry of the channels was chosen to ensure that crystallization occurs in a convection-free environment (Fig. 12a). The precipitant diffused gradually across the biomolecule solution decreasing the solubility of the biomolecule. On the right-hand side, close to the crystallant reservoir, biomolecule supersaturation was highest and induces a strong amorphous or microcrystalline precipitation. By diffusing through the channel from right to left, it created a gradient of decreasing supersaturation that results in a gradual increase of crystal size, that were located thanks to labels embossed along the channels (Fig. 12c) and grid mapping on synchrotron beamlines. The author reported on eight proteins of different sizes and sources (from bacteria to human) and an RNA oligomer crystallized in ChipX3. They adapted crystallization conditions from those initially used in vapor diffusion or batch crystallization: while the biomolecule concentration was kept unchanged, the precipitant concentration was increased by a factor of 1.5–2, as recommended by Otalora et al. (2009).

- The authors tested different materials and the last version of the chip called the ChipX3 was fabricated in cyclo-olefin-copolymer (COC) by hot embossing. It provided a stable platform for crystal storage, handling, shipment and in situ analysis by serial crystallography (de Wijn et al. 2019). The funnel-like shape of these reservoirs facilitated the contact between the biomolecule and the precipitant solutions, and avoided trapping air bubbles, which could prevent the diffusion process.

In conclusion of this part, all of these experiments with vapor-diffusion, dialysis and counter-diffusion methods, are material and time saving because they screen crystallization conditions in a single trial as shown in Fig. 2. Therefore, they partially decouple the physical mechanisms of nucleation and growth (Shim et al. 2007b, Chayen 2004). However, they generally rely on irreversible kinetic processes, which are difficult to control and optimize and they do not measure the biomolecule concentration (Shim et al. 2007b). The way to know the solute concentration and to control kinetic processes is to employ the batch method where known solutions of biomolecule and precipitant are incubated, varying the quantity of precipitant of the temperature. In microfluidics, batches are droplets of pL-nL with fixed composition that are generated and stored in a channel with two of the dimensions of sub-millimeter range.

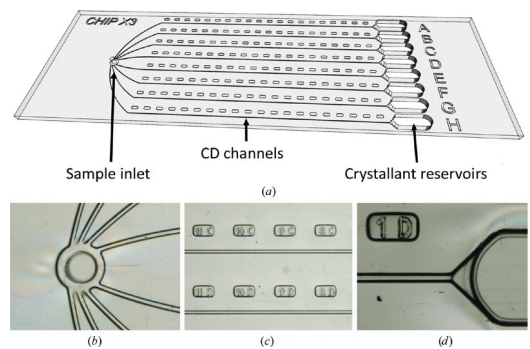

**Fig. 12** ChipX3 setup: (a) Schematic view of the chip, which has the dimensions of a microscope slide and eight channels with a straight segment of 4 cm and a cross-section of 80×80 mm. Closeup views of: (b) the inlet where the biomolecule solution, was injected into the sample inlet connecting all channels to fill simultaneously the 8 branching channels in a single manipulation, thus limiting the loading time and solution dead volumes; (c) the channels which are isolated from each other by the injection of paraffin oil before sealing with CrystalClear tape to prevent evaporation and solution movements, and labels; (d) the end of the channel with the reservoirs A to H (crystallant reservoirs) that were filled with precipitant solution before sealing them with CrystalClear tape. Reprinted with permission from Ref. (de Wijn et al. 2019). Copyright (2019) International Union of Crystallography

### II.4. Microbatch-based derived microfluidics

In batch derived methods using temperature driven crystallization (Fig. 2), microfluidics was introduced to dispense the crystallization samples in small droplets of pL-nL. These methods called droplet-based microfluidics consist in segmenting a stream of biomolecule solution (dispersed phase) with a stream of immiscible phase (continuous phase), with the assistance of syringe pumps and junctions. A tremendous advantage of droplet-based methods is their ability to change experimental parameters such as the biomolecule-to-precipitant ratio or the total sample volume by simply adjusting the flow rates of the streams. Thus, it is possible to screen different points in the phase diagram. Each droplet is considered an independent microreactor, enabling a large number of experiments to be carried out under identical conditions for statistical approach. In addition, droplets also avoid many of the problems encountered in bulk systems such as nonuniform temperatures, the influence of the reactor itself and the presence of impurities. However, the stabilization of droplets during generation, flow and incubation in the channel are proceeded using or not surfactant (detergent).

*II.4.1. Droplet-based microfluidics using surfactant*

Fraden group generated an emulsion composed of droplets of biomolecule solution in a continuous phase, using a microfluidic device fabricated in PDMS (Fig. 13a) (Akella et al. 2014). The surfactant is PTFE−PEG block copolymer, which adsorbed at the interface drop-continuous phase (Fig. 13b). In absence of surfactant droplets would coalesce. They designed this surfactant to prevent adsorption of the protein to the drop-oil interface by depletion of protein (role of PEG).

Protein to crystallize was mixed with a precipitant on-chip to avoid any nucleation before starting the experiment (Fig. 13a). The homogeneity in composition of protein and precipitant mixture in the droplets produced using a co-flow is assayed by measuring the cloud point, from which they determined that droplet-to-droplet variation in composition is less than a few percent. The emulsion droplets are then loaded in a rectangular capillary with inner dimensions of width, 1 mm (height, 50/100μm) and the ends of capillary were sealed with a mixture of equal parts of vaseline/vanoline, and low melting temperature paraffin wax. Sealing was very important because the droplets are less than 1nL in volume (#50 μm diameter), and just a minute amount of mass transport will dehydrate the drops.

The authors studied nucleation rates of lysozyme crystals based on counting the number of drops without crystals as a function of time after a quench to deep supersaturation. They measured two nucleation rates, denoted slow and fast: the slow rate varies with temperature whereas the fast one is independent of temperature. They analyzed nucleation rates within the context of Classical Nucleation Theory (CNT), which adequately describes their observations.

Furthermore, in the case of fast nucleation, they observed that nucleation always occurs on a protein aggregate and that as the crystal grows, the protein aggregate is consumed. Even the crystals that appeared at later times (≥2 h) were almost always clearly associated with protein aggregates. Because after 2h the probability that a nucleation event was due to a fast process is very low, they identify the aggregates as the source of the slow nucleation process too. Therefore, both rates are inconsistent with the process of homogeneous nucleation and are consistent with heterogeneous nucleation.

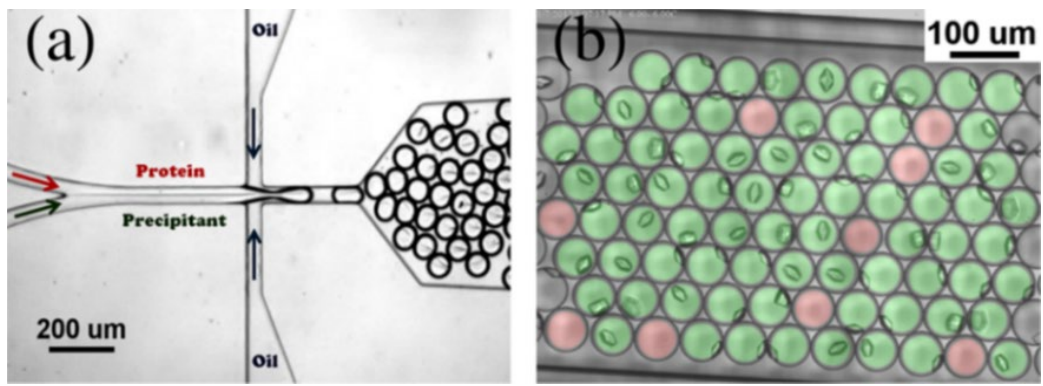

**Fig. 13** (a) Droplet generation using a co-flow microfluidic device fabricated using PDMS; (b) Detected drops with crystals highlighted in green and without crystals highlighted in red. Reprinted with permission from Ref. (Akella et al. 2014). Copyright 2014 American Chemical Society

In this method, the use of surfactant enables observing many droplets simultaneously. However, the presence of the surfactant could influence crystal nucleation and/or growth, due to its interaction with the solute.

### II.4.2. Droplet-based microfluidics without surfactant

Droplet-based microfluidics have been developed without use of surfactant in order to avoid unwanted interaction with biomolecule and/or chemical/physical influence on nucleation and growth. These devices are based on accurate control of temperature, of composition in droplets and statistical approach for crystallization fundamentals.

#### II.4.2.1. Temperature gradients on droplet-based chip

Salmon group developed a device in PDMS for two-dimensional screening of solubility, temperature vs. composition (Fig. 14). Based on thermal transfers, their microfluidic chip realized regular temperature gradients along the storage channels (Laval et al. 2007a). Each channel corresponded to a concentration ($C_1$ to $C_{10}$) as shown in Fig. 14a, and two Peltier modules placed underneath the device at positions marked by the two dotted areas, controlled the temperature field of the chip. Thin thermocouples inserted in the channels measure the temperature of the device along three series of positions parallel to the storage channels (above $C_1$, between $C_5$ and $C_6$, and below $C_{10}$). The PDMS chip was sealed on a glass slide to improve observation (Fig. 14b). With this method, the author could directly read the solubility diagram of adipic acid (see part III.1).

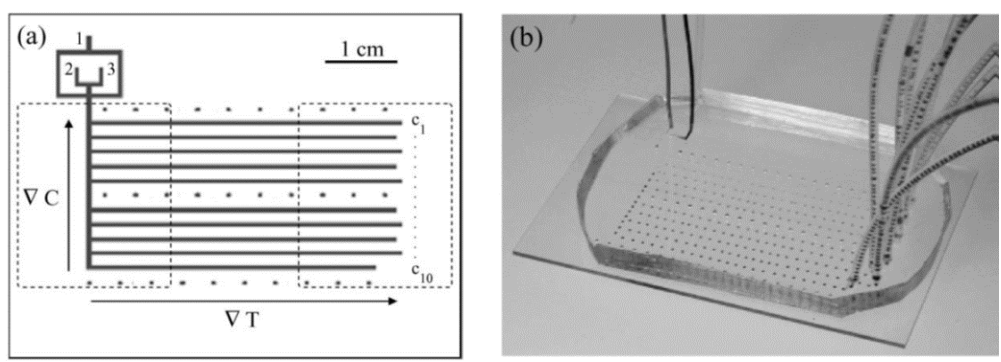

**Fig. 14** (a) Design of the microfluidic device (channels width 500 mm): silicone oil was injected in inlet 1 and aqueous solutions in inlets 2 and 3. The two dotted areas indicate the positions of the two Peltier modules used to apply temperature gradients. The three lines of dots mark the positions of temperature measurements. (b) Picture of the microfluidic chip made of PDMS sealed with a glass slide: example of droplets containing a colored dye at different concentrations stored in the ten parallel channels. Reprinted with permission from Ref. (Laval et al. 2007a). Copyright 2007 Royal Society of Chemistry

**II.4.2.2. Hybrid method for simultaneous screening and optimization of crystallization**

Ismagilov group reported a "hybrid" approach that combines screening and optimization steps, a single hybrid experiment that would eliminate the need to wait for the outcome of the initial screen before carrying out subsequent optimizations. Their device is composed of a four-inlet PDMS device coupled to Teflon capillary (Fig. 15) (Li et al. 2006).

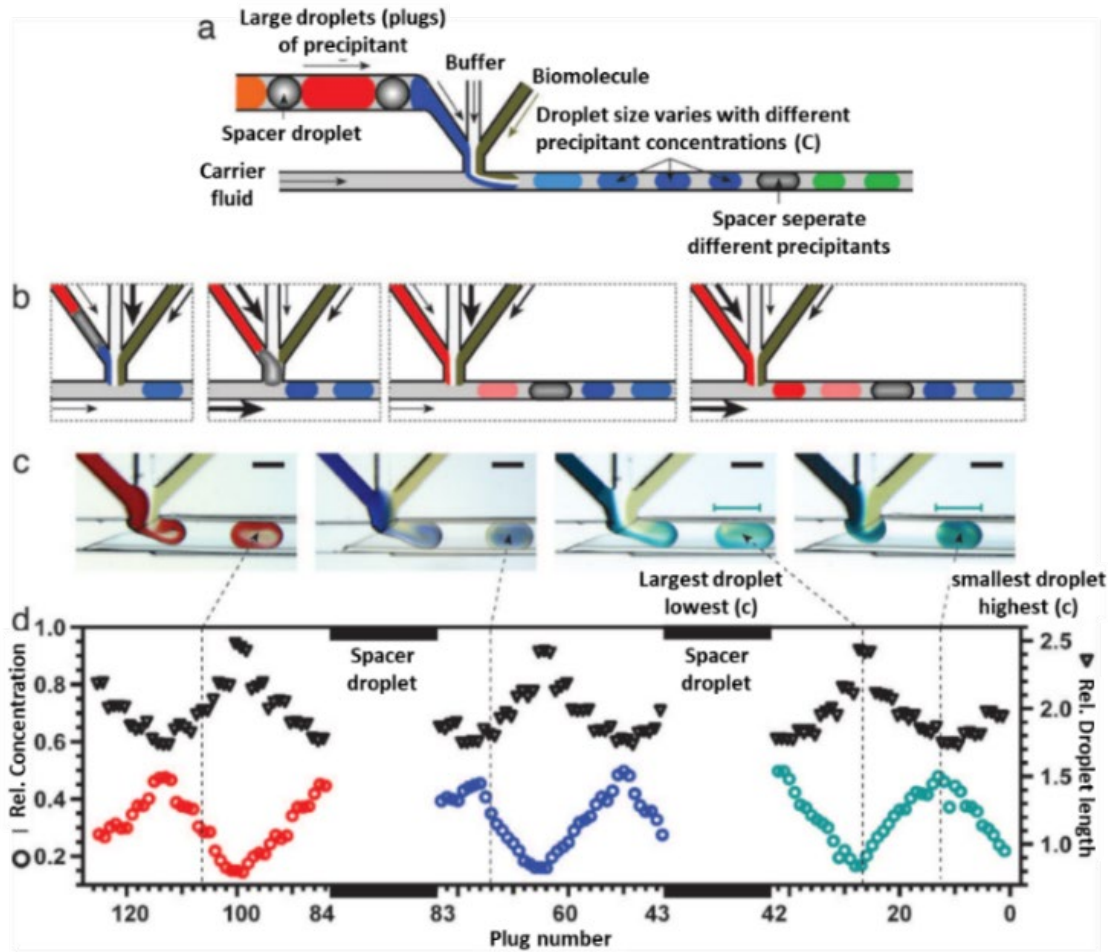

**Fig. 15** (*a*) Schematic illustration of the hybrid method for simultaneous screening and optimization of biomolecule crystallization: By one of the inlets, many distinct precipitants were sequentially introduced as ~140 nl plugs droplets into a microfluidic device, separated by spacer droplets to differentiate the precipitants. By two other inlets, a biomolecule flow and a diluting buffer flow are combined to the flow of precipitants droplets, and then distributed in droplets using a flow of continuous phase (called carrier fluid) by the last inlet. (*b*) Combination of streams in several ratios (black arrows) to form droplets containing different precipitants at different concentrations and indexing these concentrations by droplets size. (*c*) Illustration of the link between concentration and size of droplets by using droplets colored with dyes: higher concentrations correspond to smaller droplets and lower concentrations to larger droplets. (*d*) A plot quantifying the hybrid method, performed with the same precipitants in *c* but marked with fluorescent dyes instead of absorption dyes. Relative (Rel.) concentrations inside the droplets and sizes of droplets were measured from quantitative fluorescent images. Reprinted with permission from Ref. (Li et al. 2006). Copyright (2006) National Academy of Sciences, U.S.A.

To validate their hybrid method and demonstrate its applicability, authors crystallized model membrane proteins: reaction center (RC) from *R. viridis and* Porin from *R. capsulatus* (Li et al. 2006). To be solubilized in an aqueous solvent, membrane proteins need surfactants and viscous precipitants (polyethylene glycol), making crystallization conditions challenging. RC from *R. viridis*, shows a transition from slight precipitation to large single crystals or small microcrystals. Another precipitant gave a transition from precipitation to phase separation. For X-Ray measurements, crystals obtained in the hybrid device are too small. Hence, the authors scaled up the conditions determined with the hybrid screens in a silanized glass capillary of 600 µm diameter. In larger plugs, they obtained larger crystals of high quality for Porin from *R. capsulatus* that they could manipulate in plugs without any damage.

**II.4.2.3. Droplet-based microfluidics for fundamental studies**

Nucleation is a stochastic process, where random fluctuations in a solution lead to the formation of a critical nucleus. Hence a statistical approach is required (Lee 2014). which means that hundreds of droplets have to be stored in the same condition of composition and temperature. In our group, we first used droplet-based microfluidic chip which was in PDMS (Fig. 16), due to its ease for sealing without distorting channels geometry (Ferreira and Castro 2023). The basic microdevice design was fabricated by soft-lithography techniques (Laval et al. 2007b, Ildefonso et al. 2012a). In the aim to screen different temperatures with hundreds of identical elongated droplets (plugs) containing identical composition, we developed a procedure to fill several storage chips (chips 2 of Fig. 16a) with droplets generated in a plug factory (chip 1 of Fig. 16a) (Ildefonso et al. 2012a).

In the plug factory, protein and salt solutions were injected (points 2 and 3 in Fig. 16a) with syringe pumps. 100-200 droplets were generated by segmenting the stream of solution with a stream of silicone oil as continuous phase (point a in Fig. 16a). For a channel of 500 µm diameter, each plug has a volume of 200 nL with a volume polydispersity of only a few percent (Zhang et al. 2015a, Ildefonso et al. 2011). The storage chip is connected to the outlet of the plug factory zone (point b in Fig. 16a) using micro-tubing in Teflon (Ildefonso et al. 2012a, Hammadi et al. 2013).

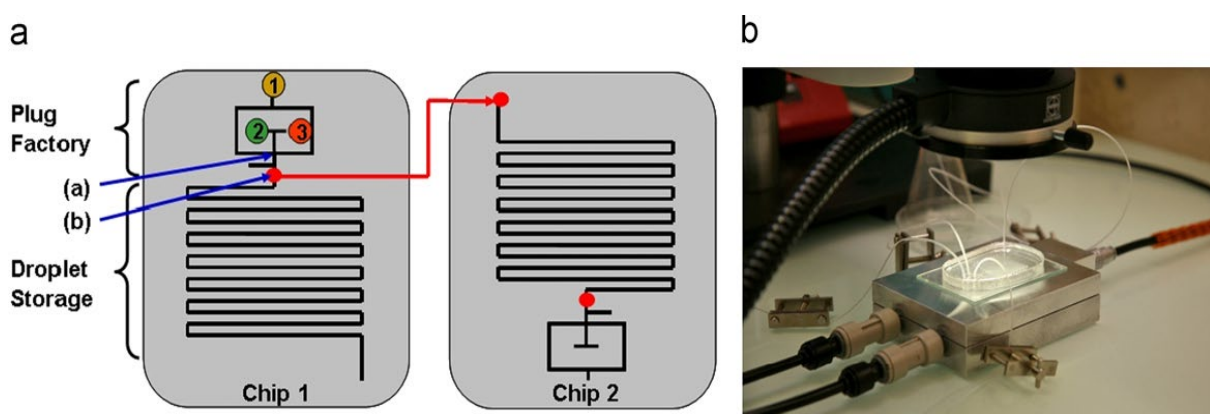

**Fig. 16** Droplet-based microfluidic system in PDMS for statistical approach: a) Design of the microdevice with 2 chips- chip 1 for the plug factory and chip 2 for droplet storage (channel width 500 µm) and b) microfluidic experimental set-up with thermostatted sample-holder. Reprinted with permission from Ref. (Ildefonso et al. 2012a). Copyright (2012) Elsevier

We crystallized lysozyme in aqueous solution by cooling the droplets to 6 °C. By observing hundreds of droplets, we illustrated the stochastic behavior of nucleation. We obtained large crystals (#100 µm) and we even isolated a metastable phase (Ildefonso et al. 2012a). We also studied phase diagram of lysozyme and nucleation kinetics, resulting in the determination of the effective interfacial energy crystal/solution (see part III.3.) (Ildefonso et al. 2011, 2012a, 2012b, 2013, Candoni et al. 2012a, 2012b).

This device has proved to be an easy-to-use microfluidic tool for biomolecules soluble in aqueous solutions. However, if the geometry needs to be modified, to integrate inlets for example, a new mold has to be created by soft-lithographic techniques, which is costly and time-consuming. Moreover, the number of experiments is limited due to the porosity of PDMS, and its use is often even unique (non-recyclable). Similarly, droplets can only be stored for a maximum of 24 hours, since for longer storage times, evaporation due to the permeability of PDMS is no longer negligible (Laval et al. 2007a, Ildefonso et al. 2011). Moreover, PDMS is not compatible with organic solvents such as acetone.

In order to make the microfluidic system easy to fabricate without using soft-lithography techniques, we developed a microfluidic circuit composed of HPLC consumable material: junctions and fittings are in polyether ether ketone (PEEK) and tubing in Teflon (PFA or PFE), inner diameters from 150 µm to 1 mm are available corresponding to droplet volume range from nL to µL (Zhang et al. 2015b). These polymers are compatible with almost all solvents, showing excellent resistance and no solvent evaporation, which enables universal use of the circuit, whatever the solvent, at a wide range of temperatures and without evaporation (Ildefonso et al. 2013).

For a statistical approach of crystallization, we stored sets of hundreds of droplets at the same temperature but varying condition of biomolecules or precipitant concentrations. For saving material and time, we developed methods to fill directly the tubing for screening and optimization experiments (Gerard et al. 2018) (see part II.5) and to prepare saturated solution directly from powder in the set-up (see part III) for studies on phase diagram and polymorphism (Peybernès et al. 2018, Peybernès et al. 2020, Lambert et al. 2023). The entire platform (Fig. 17) enabled to study a biomolecule of interest from the powder (Fig. 17b) to the crystallization (Fig. 17d), using image acquisition (Fig. 17e) and spectroscopic characterization by UV-visible (Fig. 17c) and Raman (Fig. 17f) (Lambert et al. 2023). In addition, independent modules for mixing of liquids, detection of droplets and their concentration, and storage are combined (Ildefonso et al. 2012b, Zhang et al. 2017, Ildefonso et al. 2013, Gerard et al. 2018), rendering the platform extremely flexible.

We investigated crystallization of various biomolecules in organic and aqueous solvents: lysozyme in aqueous and viscous solutions (Ildefonso et al. 2011, 2012a, 2012b, 2013, Zhang et al. 2017, Gerard et al. 2018); rasburicase (Zhang et al. 2015b) and human quinone reductase 2 (QR2) (Gerard et al. 2017, Gerard et al. 2018) as biomolecules. Screening and optimization of crystallizations and research of polymorphs will be presented in part III, as well as X-Ray data collection. As the device is composed of commercially available components that were plug-and-play, it is easy to use for non-specialists in microfluidics. We easily transferred it in standard laboratory environments, for instance in industries (Sanofi and Servier) (Peybernès 2019, Gerard 2017).

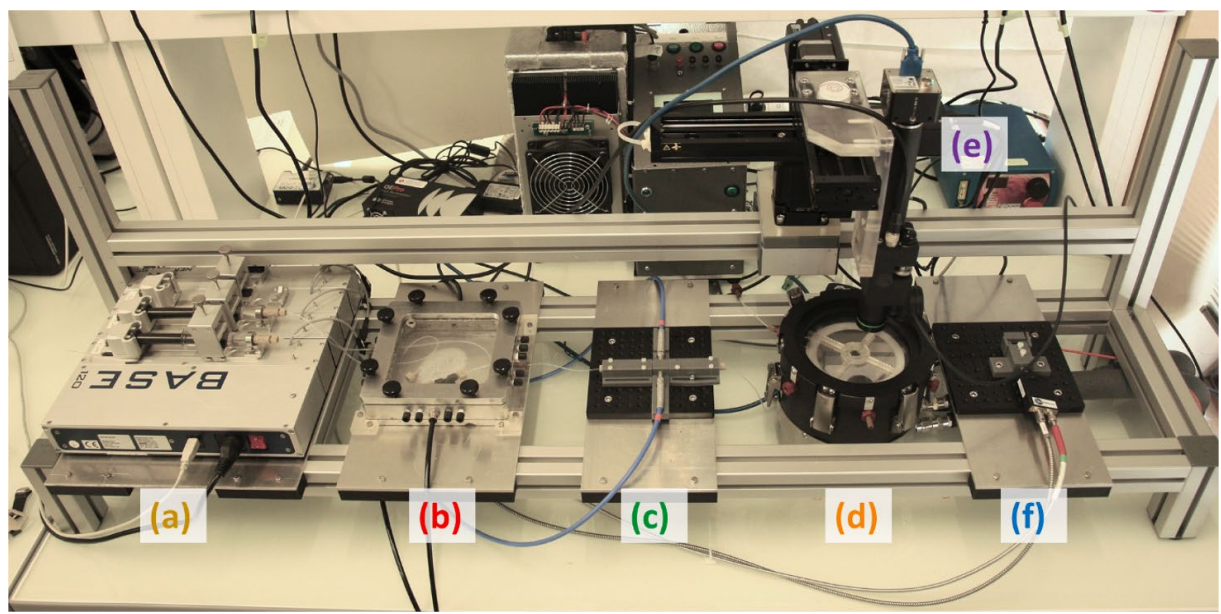

**Fig. 17** Microfluidic platform: (a) Syringes and pumps; (b) Solutions and droplet generation with temperature control; (c) UV characterization; (d) Droplet storage and cooling; (e) Optical characterization; (f) Raman characterization. Reprinted with permission from Ref. (Lambert et al. 2023). Copyright (2023) Elsevier

To promote easy-to-use microfluidic tools for screening and optimizing crystallization conditions, improvements have been made in terms of parallelization of experiments and mixing of aqueous solution and organic solvent, while saving time and material.

**II.5. Parallelization of experiments**

In the microfluidic tools described above, precipitants are introduced either by pipetting and loading directly in wells/reservoir or into the inlet of the fluidic path or by injection with pumps (syringe or pressure). Even in the method of Ismagilov group, where precipitants are in the form of plugs (Fig. 15), syringes have been used to generate these plugs. Hence, all of these methods test multiple precipitants by refilling manually wells/reservoir or injecting with multiple syringes or pumps and/or valves, leading to material and time-consuming experiments.

• A way to reduce the volume of precipitants consists in filling directly capillary with large plugs with the needed volume of each precipitant (Fig. 18a and b) (Gerard et al. 2018).

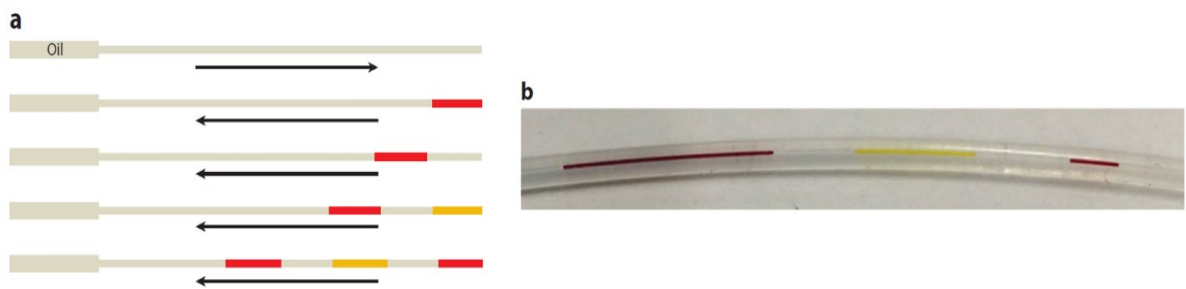

**Fig. 18** Filling directly capillary of 150 μm inner diameter by sucking aqueous solutions with dyes. (*a*) Schematic illustration of the tubing first filled with the continuous flow (in grey) and successively refilled with red and yellow dyes solutions, by interspacing continuous phase between solutions. (*b*) Picture of capillary with plugs of dye, first of red (0.22 μL), then yellow (0.13 μL), and then red (0.05 μL), separated from each other by a continuous phase spacer to prevent plugs from mixing. Arrows indicate the flow direction. Reprinted with permission from Ref. (Gerard et al. 2018). Copyright 2018 American Chemical Society

• Trivedi et al. reduced the number of syringes by mixing all of the precipitants in one syringe and separating them with a chromatography column to generate droplets of each precipitant (Fig. 19) (Trivedi et al. 2010). However, it requires to use the same solvent to solubilize all of the precipitants. Compare to Trivedi et al., our system of tubing direct filling enables the use of a wide range of solvents.

Then precipitants solutions are added to biomolecule solution by mixing directly in a multiport junction or incorporating into droplets of biomolecule solution through a junction. These techniques improve the flexibility of microfluidic methods.

**Fig. 19** Separation of precipitants from a mix of precipitants with chromatography separation column and absorbance detectors for each precipitant. Reprinted with permission from Ref. (Trivedi et al. 2010). Copyright 2010 Royal Society of Chemistry

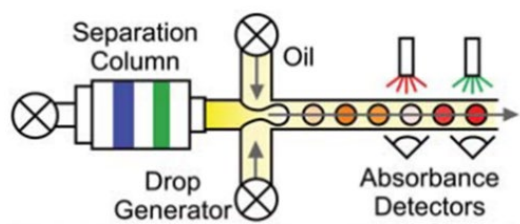

In conclusion, all of the microfluidic systems described above reached to obtain crystals in a reasonable time due to sufficient supersaturation obtained by concentrating biomolecule and/or precipitant by vapor-diffusion, dialysis methods or counter-diffusion methods, and/or cooling biomolecule solution. Some of these methods provide a statistical approach for investigating fundamental properties of crystallization.

## III. Fundamental properties of crystallization

Microfluidic devices are used for a miniaturization and parallelization of experiments to explore fundamental properties of crystallization such as nucleation kinetic, nucleation frequency and interfacial energy crystal/solution. For crystallization of biomolecules, the phase diagram serves to fix conditions of experiments in terms of concentration and temperature, depending on the medium of crystallization (solvent with or without crystallization agents) (Fig. 2). In the phase diagram, the key parameter is the thermodynamic solubility $C_s$ defined as the concentration for which crystals and solution are at equilibrium. In order to observe nucleation or growth, this solution must be supersaturated. The supersaturation $\beta$ is defined by the ratio between the concentration C and Cs and it is the driving force for nucleation and growth. According to the value of β, two zones correspond to nucleation or growth promotion, respectively. We will show how the thermodynamic solubility curve and the limit between nucleation and growth zones are explored with microfluidic systems.

### III.1. The thermodynamic solubility

The thermodynamic solubility is measured through equilibration of a biomolecule solution with a crystalline solid phase at a given temperature T (Detoisien et al. 2009, Pusey and Gernert 1988, Broutin et al. 1995).Two methodologies are reported to determine the thermodynamic solubility at a given T either by dissolving crystals into an undersaturated solution or by exposing crystals to a supersaturated solution leading to crystal growth (Trevino et al. 2008, Ferreira et al. 2020, van Driessche dt al. 2009, Lau et al. 2007) (see chapter 2 of part A of the present book). In both methodologies, at the end of the experiment, the system is brought at equilibrium and crystals coexist with a saturated solution. However, the second methodology requires that crystallization conditions of the target biomolecule to be known, and so it is limited to model proteins (or well-known proteins). Moreover, measurements can be laborious and time consuming (Ferreira and Castro 2023). Therefore, the method of dissolution is more applicable and the experiment is repeated at different temperatures to plot the solubility curve.

• Salmon group measured directly the solubility diagram with their device described in Fig. 14, by storing hundreds of droplets (#100 nL) of various chemical compositions in parallel microchannels and applying large temperature gradients (Laval et al. 2007a). Fig. 20 shows ten values of solubility temperature obtained in a single chip at concentrations $C_1$ to $C_{10}$, enabling to draw the solubility curve of adipic acid. For a given concentration, the solubility temperature is just above the minimum temperature at which the droplet was crystallized (appearing white). By measuring directly solubility diagrams with this device, the authors reduced the time of experiments to only 1 hour.

**Fig. 20** Direct reading of the solubility diagram of adipic acid: the dotted line bounding droplets containing crystals gives an estimation of the solubility limit. Reprinted with permission from Ref. (Laval et al. 2007a). Copyright 2007 Royal Society of Chemistry

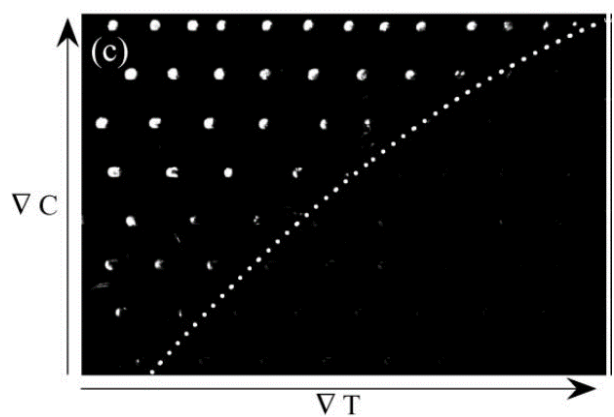

• By using microfluidic system, the total volume of solution in droplets is reduced to μL. However, the use of syringe requires a volume of stock solution to be prepared of at least hundreds of μL. On the one hand, the solution may crystallize in the syringe, and on the other, some biomolecules may only be available in small quantities. In the platform of Fig. 17, we generated saturated solution directly from powder by flowing a solvent through a bed of powder contained in a UHPLC column (Fig. 21a) (Peybernès et al. 2018).

At the outlet of the filter, due to the biomolecule dissolving, the solution concentration is at saturation Cs. This saturated solution is then analyzed online by a UV-Vis spectrometer and the Beer-Lambert's law is used to determine Cs from the measure of absorbance. However, two limitations appear: 1) Cs must be in the linear part of the Beer-Lambert's law and 2) the absorbance must be calibrated for each molecule, requiring additional material. To overcome these limitations, we replaced the calibration step by only one measure of the absorbance on a standard solution of 1 mg/mL. This low concentration guarantees the location in the linear part of the Beer-Lambert's law and to be lower than Cs. Hence, the outlet solution at Cs is diluted to reach 1mg/mL and knowing the volume of dilution Cs is calculated. Therefore, the volume of solvent used for dilution is calculated from the flow rate programmed through a syringe pump (Fig. 21b).

Measures of the entire solubility curve at a variety of temperatures ranging from 20 °C to 60 °C is realizable in just few hours, consuming small quantities of material (3–30 mg), with a concentration range from one to hundreds of milligrams per milliliter.

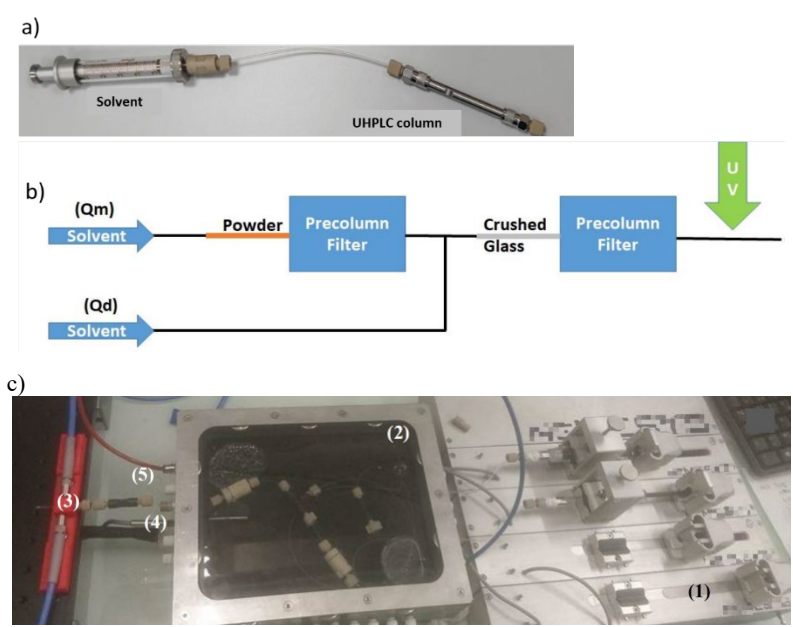

- **Fig. 21** (a) Scheme of powder dissolution to generate solution at saturation with syringe of solvent, a tubing and a UHPLC column. (b) Microfluidic platform for the measure of Cs by dilution and absorbance measurement: (1) programmable syringe pump; (2) incubator filled with thermoregulated water to control the temperature T; (3) on-line analyze of the concentration in the flow of solution by a UV-Vis spectrometer; (4) temperature probe and (5) outside loop insuring solution thermal equilibrium to room temperature. Reprinted with permission from Ref. (Peybernès et al. 2018). Copyright 2018 American Chemical Society

### III.2. The zones of metastability and supersolubility

According to Classical Nucleation Theory (CNT), from the moment solution is supersaturated, nucleation can occur. However, a certain time elapses before the appearance of the first crystal and its detection. The Metastable Zone (MZ) is the zone in which the supersaturated solution can stay for a given time, without losing its metastability (Kashchiev et al. 1991). It displays only crystal growth but no nucleation. For temperature much lower, the solution is highly supersaturated and a huge number of crystals can nucleate instantaneously leading to a precipitation but large single crystals are not likely to form from this precipitate. This zone is called the supersolubility zone (Sommer and Larsen 2005) or zone of spontaneous nucleation. The limit between the metastable and the supersolubility zones is called the metastable zone limit (MZL). MZL is a kinetic curve drawn by connecting the minimum temperature of the MZ (TMZ) for each concentration explored during a given time.

#### III.2.1. Isothermal and polythermal experiments

To determine MZL, two kinds of experiments are possible, isothermal and polythermal, which are graphically illustrated in Fig. 22 (Devos et al. 2021). For polythermal experiments, a solution at a given concentration is cooled at a constant cooling rate, until crystals are detected. For isothermal experiments, the temperature is kept constant throughout the experiment at a given value. The time from the start of the experiment until the appearance of crystals is recorded. Isothermal and polythermal experiments generally require similar setups and time.

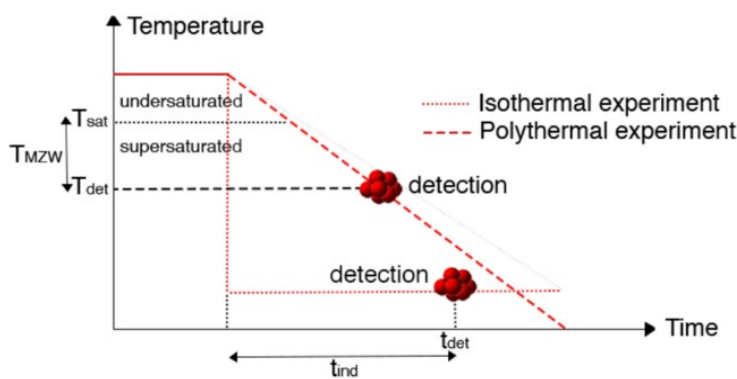

**Fig. 22** Graphical representation of isothermal and polythermal experiments for the determination of metastable Zone Limit cooling crystallization. Adapted with permission from Ref. (Devos et al. 2021). Copyright 2021 American Chemical Society

As in polythermal experiments the MZ is strongly dependent on the control of the cooling rate, isothermal experiments are easier to proceed for the determination of TMZ. Using the microfluidic system described in Fig. 16, we stored nL droplets at different temperatures $T$ for the same time $t$, that is chosen long enough for the crystals to nucleate. During $t$, if $T < T_{MZ}$, crystals nucleate, and if $T > T_{MZ}$, no crystals appear. Then we stored the droplets at a temperature just below the temperature of solubility, to make crystals grow, as growth is known to be optimal in the MZ. As droplets were of small volume, the number of crystals was low and could be counted for each temperature. This measurement was performed for several concentrations, giving several points of the metastable zone limit (MZL). Each point corresponded to a statistical treatment on 100 droplets. For a given concentration, the lowest temperature at which no crystals are observed, is the corresponding $T_{MZ}$. So $T_{MZ}$ was extrapolated (interpolation) from a plot of the number of crystals in the solution versus temperature. We evaluated MZ for 20 hours of storing and we drew the diagram phase for lysozyme in an aqueous solution (Fig. 23a) (Ildefonso et al. 2011) and for caffeine in ethanol (Fig. 23b) (Ildefonso et al. 2012b).

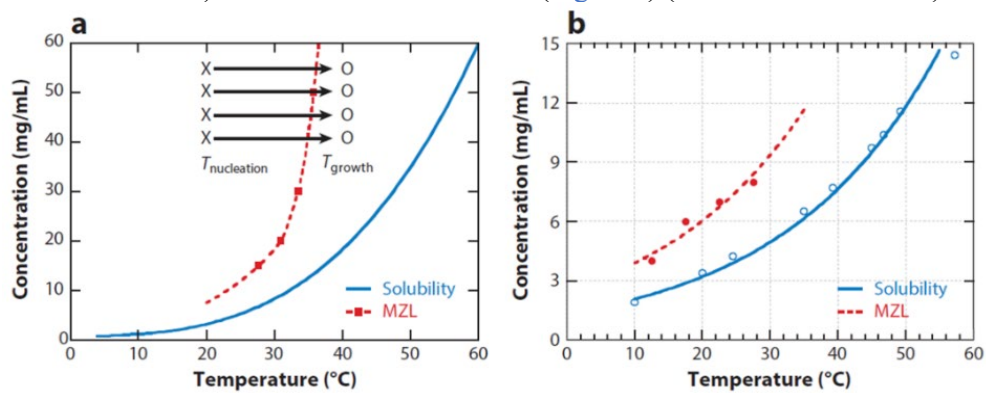

**Fig. 23** Phase diagram of concentration versus temperature obtained on the device of figure D, with the solubility curve (*solid line*) and the metastable zone limit (MZL) (*dashed line*): (*a*) for lysozyme in an aqueous solution and (*b*) for caffeine in ethanol (lines improve legibility). Figure adapted with permission from from Ref. (Ildefonso et al. 2011). Copyright 2011 American Chemical Society. And from Ref. (Ildefonso et al. 2012b). Copyright 2012 American Chemical Society

The same curves can be drawn for other physical or chemical parameters such as salt concentration, polymer concentration or pH. For instance, we determined the phase diagram of quinone reductase (QR2) in ammonium sulfate. In Fig. 24A, optimal conditions leading to high-quality crystals were found testing at least 60 droplets of 2 nl per experimental condition. The crystals obtained for each experimental condition (pointed in Fig. 24A) are shown in Fig. 24B with 5 representative droplets for each experimental condition: (a) just saturated, (b) and (c) with crystals grown in the MZ, (d) and (e) with precipitates nucleated in the supersolubility zone (Gerard et al. 2017).

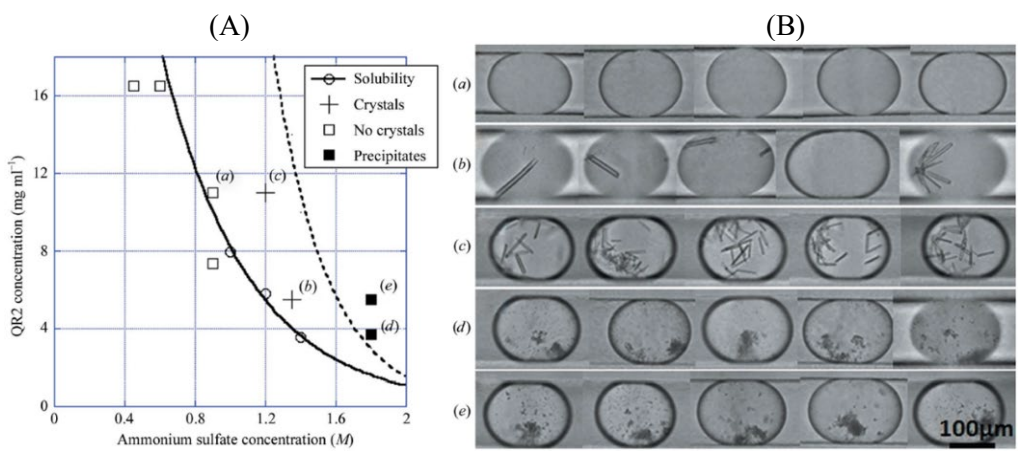

**Fig. 24** Phase diagram of quinone reductase (QR2) in ammonium sulfate at pH 8 (20 mM Tris–HCl, 150 mM NaCl, using a device similar to figure M: (A) plotted from 8 experimental conditions (the dashed line is a guide for the eye to separate the crystallization and precipitation zones and error bars are indicated by the size of the markers); (B) photographs of five representative droplets obtained as 2 nl droplets in a Teflon capillary (150 mm inner diameter) after 24 h. Reprinted with permission from Ref. (Gerard et al. 2017). Copyright (2017) International Union of Crystallography

### *III.2.2. Influence of operating parameters on the MZL*

In contrast to the solubility curve, the metastable limit is not thermodynamically defined and strongly depends on process parameters such as cooling rate, impurities (Nyvlt et al. 1985) and volume (Hammadi et al. 2013):

- The influence of cooling rate is due to: if the cooling rate is high, nuclei appear at a lower temperature that the case for low cooling rate is reverse.

- The influence of impurities is due to their role act as effective nucleating agents, because of the reduction of the energy barrier to nucleation. When a solution is divided into small droplets, the impurities present are distributed over the volumes (Qu et al. 2006). The opposite phenomenon is observed with evaporative crystallization of small sample volumes of D-mannitol, as a result of the increase of impuritities concentration (Buanz et al. 2019). This effect of volume is confirmed by observation of MZL increase with filtered solutions compared to unfiltered ones, and so crystals were longer to nucleate (Mohd et al. 2020).

- The influence of volume was explored by nucleating isonicotinamide in ethanol in 3 different volumes, i.e. 250 nl in our microfluidic device of Fig. 16 and 250 µL and 1 mL in using a multi-well set-up (Detoisien et al. 2010). In 250 nL, we observed no nucleation. The other volumes led to crystals (Fig. 25b) and the MZ is represented in Fig. 25a. It is decreased of 6.5 °C when the volume decreases from 1 mL to 250 µL (volume divided by 4). Consequently, for a volume of 250 nL (volume divided by $10^3$), the MZL would be in the negative range of temperature, explaining the absence of nucleation. This effect of volume on nucleation is discussed in part IV.

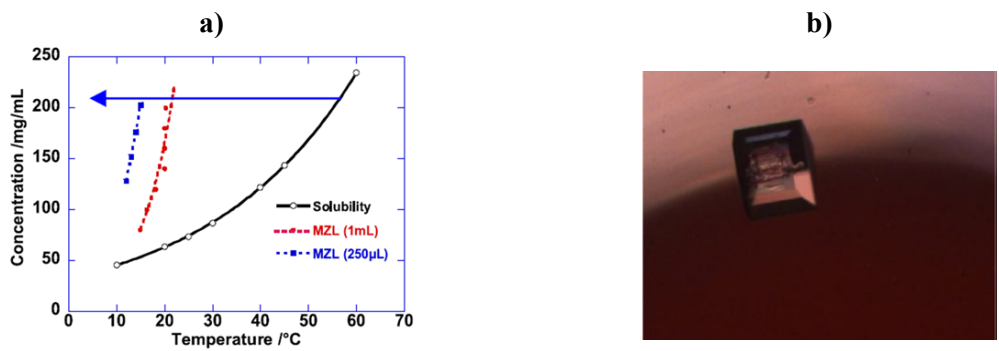

**Fig. 25** Crystallization of isonicotinamide in ethanol in 2 different volumes in multiwel: (a) MZL for 250 µL and 1 mL; (b) crystal. Reprinted with permission from Ref. (Hammadi et al. 2013). Copyright (2013) Elsevier

### III.3. Nucleation kinetics

The MZL defines two zones where nucleation and growth are kinetically or thermodynamically promoted, respectively. The nuclei appearing in the supersolubility zone leads to a depletion of the supersaturation preventing crystal growth. To solve this problem, after nucleation occurs, the temperature is increased (into the MZ), such that the existing nuclei can grow to a detectable size without nucleation of other crystals. Galkin and Vekilov (1999) propose to put the temperature of the growth zone close to the solubility curve, to suppress additional nucleation. As described in part II, this decoupling of nucleation and crystal growth is possible in microdialysis methods by changing solvent conditions (Shim et al. 2007b). and in free-interface diffusion (FID) techniques due to precise and reversible control over the crystallization conditions combined with temperature control (Junius et al 2020).

### *III.3.1. Double pulsed method*

Decoupling nucleation and growth of crystals is used in the double pulse method [74] to determine the nucleation rate measurements. In our group, we tested double pulsed method with the droplet-based device of Fig. 16, by imposing a sequence of temperatures for which we used the phase diagram (solubility curve and MZ as in Fig. 23a):

- Droplets are first generated at a temperature chosen to prevent crystal nucleation, $T_{growth}$.
- Then the temperature is lowered to $T_{nucleation}$ to obtain crystal nucleation.
- After a period of $\Delta t$ (nucleation time), temperature is raised from $T_{nucleation}$ to $T_{growth}$. At $T_{growth}$, supersaturation is at a level where there is no nucleation, but the crystals already formed can grow.

The precautions to be taken with this method are as follows:

- $T_{growth}$ and $T_{nucleation}$ are chosen to prevent or obtain crystal nucleation, respectively.

- $\Delta t$ is chosen to guarantee that solute concentration was constant in the droplets, during the step of nuvleation (Bacchin et al. 2022).

- As the critical size is inversely proportional to the temperature difference between the temperature of the solution and the solubility temperature, the critical size in the growth zone is higher than in the nucleation zone (Dixit et al. 2001). Hence, crystals nucleating in the nucleation zone, may dissolve in the growth zone.

- Although the chance of nucleation is minimal in the growth zone, it is necessary to confirm that the crystals identified in the growth zone have all nucleated in the nucleation zone.

- In addition, heterogeneous or secondary nucleation can still occur due to the considerable level of supersaturation (Chauhan et al. 2023).

### *III.3.2. Nucleation rate*

The nucleation rate $J$ is the number of crystals that form in the supersaturated solution per unit of volume and per unit of time. $J$ can be determined using the double-pulse technique (DPT) to separate nucleation and growth (Galkin and Vekilov 1999, Revalor et al. 2010, Nappo et al. 2018). After the sequence described above, the crystals nucleated at $T_{nucleation}$ during $\Delta t$ are counted after the growth stage. Hence crystal distribution is fitted with a Poisson law as previously reported by Galkin & Vekilov (1999), leading to the average number of crystals nucleated $N$ in one droplet versus nucleation time $\Delta t$.

For lysozyme explored with device of Fig. 16, $N$ varies linearly with time and the fit passes through the origin (Fig. 26a) (Ildefonso et al. 2011). Then, the slope of this fit corresponds to the nucleation rate $J$ for a given supersaturation. The evolution of $J$ versus supersaturation shown in Fig. 26b is in good agreement with the data of Galkin & Vekilov (1999, 2000) obtained for the same composition of lysozyme solution at similar temperature, using larger droplets (1 µL) suspended in Teflon wells.

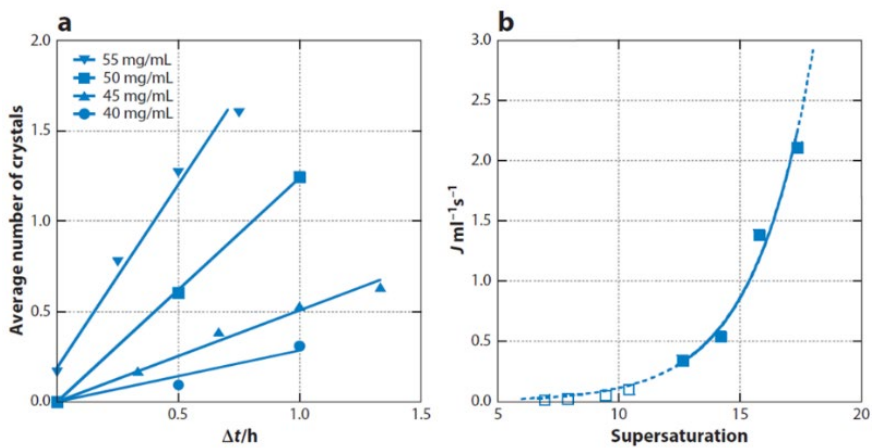

**Fig. 26** Double-pulse technique (DPT) to determine the nucleation rate $J$ for lysozyme with the device of Fig. 16: *(a)* Crystal distribution fitted with a Poisson law (Galkin and Vekilov 1999), leading to the average number of crystals nucleated $N$ in one droplet versus nucleation time $\Delta t$. (b) Nucleation rate $J$ versus supersaturation, at 20 °C, from the slope of the straight lines of *(a)*; *(open squares)* data obtained at 12.6 °C from Galkin & Vekilov (1999). The deviation is within the symbol between 1 % and 3 % of deviation on average number of crystals (for details on the estimation of error, see the paper from Galkin & Vekilov (1999). Reprinted with permission from Ref. (Ildefonso et al. 2011). Copyright 2011 American Chemical Society

Moreover, the supersaturation range experimentally accessible when the nucleation volume is reduced to the nanoliter-to-picoliter range is increased for kinetic (Candoni et al. 2015) and thermodynamic (Galkin and Vekilov 1999, Grossier and Veesler 2009) reasons. Straight lines confirm steady-state nucleation and their passage through the axis origins, except for the highest supersaturation (Ildefonso et al. 2011), indicates that the data were not affected by major heterogeneous nucleation (Galkin and Vekilov 1999).

The presence of crystals at $\Delta t = 0$ for a lysozyme concentration of 55 mg/mL indicates that this solution is near its limit of metastability and nucleation probably occurs during transfer to the incubator, after filling. However, these crystals had no influence on the slope obtained over time.

### III.3.3. Effective interfacial energy crystal/solution

In the case of crystals appearing in droplets containing no crystals of that phase, the nucleation is called primary nucleation. In addition, primary nucleation can occur through two mechanisms, i.e., homogeneous nucleation, where nuclei form spontaneously in the bulk solution, or heterogeneous nucleation due to foreign molecules or particles, bubbles, crystallizer walls, liquid−liquid or liquid−air interfaces (see chapter 3 of part A of the present book). The surface of these foreign components can be a place on which nucleation can preferentially occur, thus acting as a nucleation catalyst (Turnbull 1952). Hence, two kinds of nucleation are distinguished, homogeneous and heterogeneous with their own interfacial energy.

#### III.3.3.1. Interfacial energy of homogeneous nucleation

In the ideal case of absence of heterogeneity, or at least in the absence of their interaction with nucleation, the nucleation is called homogeneous. In the case of Classical Nucleation Theory (CNT) for Homogeneous Nucleation (HON) (Zettlemoyer 1969, Abraham 1974, Kashchiev 2000), the nucleation rate $J$ (number of crystals·s$^{-1}$·m$^{-3}$) can be expressed (Boistelle 1986) as:

$$J = K_0 exp\left(\frac{\Delta G^*_{HON}}{kT}\right) \quad (1)$$

where $K_0$ is a kinetic factor (m$^{-3}$·s$^{-1}$), $\Delta G^*_{HON}$ the activation free energy for homogeneous nucleation (J), $k$ the Boltzmann constant (J·K$^{-1}$), and $T$ the temperature (K). The CNT assumes a spherical form for the critical nucleus; this point was mentioned by Fletcher (1958) and is in good agreement with our observations of crystals just after nucleation (Hammadi et al. 2013). In this ideal case, an isotropic interfacial energy, $\gamma$ (J·m$^{-2}$), of the critical nucleus is used, and equation 1 becomes:

$$J = K_0 exp\left(-\frac{16\pi}{3}\frac{\Omega^2\gamma^3}{(kT)^3 ln^2\beta}\right) \qquad (2)$$

Where $\Omega$ is the volume of one molecule in the critical nucleus (m$^3$).

**III.3.3.2. Interfacial energy of heterogeneous nucleation**

In the case of presence of heterogeneities that interact with nucleation, it is known as heterogeneous nucleation (HEN) in the literature (Kashchiev 2000, Boistelle 1986, Liu 2000, Kashchiev 2003). The main idea is that the foreign substance decreases the activation free energy required to form the critical cluster, $\Delta G^*_{HEN}$. This decrease is determined by the ratio between the volumes of the cluster onto the foreign substance and the corresponding homogeneously formed spherical cluster. Introducing a factor $f(\theta)$ with $\theta$ the contact angle of the cluster onto the foreign substance, $J$ becomes:

$$J = K_0 exp\left(\frac{\Delta G^*_{HEN}}{kT}\right) \qquad (3)$$

With

$$\Delta G^*_{HEN} = f(\theta) \times \Delta G^*_{HON} \qquad (4)$$

Assuming a constant shape for the cluster (here spherical) and depending on the affinity of the cluster to the foreign substance, as characterized by a contact angle, defined from 0 to $\pi$, we deduced:

$$0 < f(\theta) < 1 \qquad (5)$$

The factor $f(\theta)$ was generalized to different shapes for clusters and foreign substances (Fletcher 1958, Liu 2000, Kashchiev 2003) and it represents the thermodynamic part of the catalytic effect of the foreign substance on nucleation. In practice, an effective interfacial energy $\gamma_{ef}$ was introduced depending on $f(\theta)$ (Kashchiev 2000):

$$\gamma_{ef} = f(\theta)^{1/3} \times \gamma \qquad (6)$$

We see that the lower $f(\theta)$ or $\gamma_{ef}$, the greater the affinity of the cluster to the foreign substance and $J$ is given by the equation:

$$J = K_0 exp\left(-\frac{16\pi}{3}\frac{\Omega^2\gamma_{ef}^3}{(kT)^3 ln^2\beta}\right) \qquad (7)$$

To determine $\gamma_{ef}$, we evaluated $J$ for lysozyme in the device in PDMS of Fig. 16 (Ildefonso et al. 2012a) and in device in PEEK/Teflon of Fig. 17 (Ildefonso et al. 2012b) generating droplets in silicone oil and fluorinated oil, respectively. In Fig. 26, $J$ varies exponentially with $\left(\frac{1}{ln^2(\beta)}\right)$ (Ildefonso et al. 2013). The value of $\gamma_{ef}$ was compared to values obtained by Fraden group (Du et al. 2009) in PDMS microfluidic device (Fig. 4) with fluorinated oil, and Vekilov group (1999, 2000) with droplets suspended in silicone oil in Teflon wells (sub millimetric) in a supersaturation range of 5−10. Table 3 summarizes values of $\gamma_{ef}$ and experimental conditions (Ildefonso et al. 2013). According to Table 3, data from different groups with the same continuous phase gave the same $\gamma_{ef}$ indicating that the interface between continuous phase and protein solution represents the main heterogeneity for lysozyme nucleation, that is to say, the "foreign substance" for HEN and not the type of device. This is an experimental confirmation of the efficiency of oil for avoiding contact between crystallizing solution and device walls by creating a "containerless" environment as pointed out by Chayen.[91] Moreover, $\gamma_{ef}$ is higher in fluorinated oil than in silicone oil, pointing to a better catalytic effect on nucleation using silicone oil than using fluorinated oil. This result shows the importance of the type of continuous phase used in microfluidic systems. However, there are too few foreign molecules, particles or bubbles in droplets to act as heterogeneities, probably due to their dilution as the solution is distributed in nanovolume droplets.

**Table 3.** Effective interfacial energy $\gamma_{ef}$ determined for different chemical natures of devices and continuous phase, by different groups. (the deviation in absolute is given in parentheses, errors are not available in (Selimpovic et al. 2009, Galkin and Vekilov 1999, Galkin and Vekilov 2000). Adapted with permission from Ref. (Ildefonso et al. 2013). Copyright 2013 American Chemical Society.

|  | Our device | Vekilov group | Our device | Fraden group |
| --- | --- | --- | --- | --- |
| Design of device | Fig. 16 | Sub millimetric wells | Fig. 17 | Fig. 4 |
| Reference | (Ildefonso et al. 2012a) | (Galkin and Vekilov 1999, Galkin and Vekilov 2000) | (Ildefonso et al. 2012b) | (Selimpovic et al. 2009) |
| Material of device | PDMS | Teflon | PEEK/Teflon | PDMS |
| Continuous phase | Silicone oil | Silicone oil | Fluorinated oil | Fluorinated oil |
| $\gamma_{ef}$ (mJ.m$^{-2}$) | 0.62 (± 0.13) | 0.56 | 0.88 (± 0.05) | 0.91 |
| Droplets volume (nL) | 250 | 1000 | 100 | 1-2 |
| $\beta$ range | 10-15 | 5-10 | 12-20 | 24−55 |

Furthermore, compared with the µL and nL volumes used by Vekilov group and Fraden group respectively, we obtained similar results with droplets of 100-250 nL. Thus, the depletion of nanovolume solution during nucleation has negligible influence on the total concentration of the solution avoiding thermodynamic effect of confinement as observed from picoliter range down (Grossier and Veesler 2009). This is in agreement with experimental results showing effects under nanoscopic confinement (Ha et al. 2004, Beiner et al. 2007, Kim et al. 2009, Stephens et al. 2010, Grossier et al. 2012). However, we observe kinetic effects as the supersaturation range experimentally accessible increases with the volume decrease.

## IV. Kinetic effect of confinement in nanovolumes

Given the stochastic nature of nucleation, the average nucleation induction time for a single nucleus $t_N$ is inversely proportional to the volume of the crystallizer $V$ and to nucleation rate $J$, for a given experimental condition. For decreasing $V$, $J$ is the same in the case of homogeneous nucleation. If the nucleation is heterogeneous, $J$ decreases because of less impurities in small volumes (Kashchiev 2000). Hence, nucleation is inhibited and higher supersaturation is accessible in the solution before the solution nucleates (Fig. 26b) (Langer and Offermann 1982, Galkin and Vekilov 2000, Pons-Siepermann and Myerson 2018).

High values of $\beta$ facilitate the appearance of new crystal phases (Hilden et al. 2003), probably because the differences between their free energies become indistinguishable in small volumes, blurring the selectivity for phases (Fellah et al. 2024). Furthermore, Ostwald's stages rule postulates that when a substance crystallizes from a high-energy non-equilibrium state, it will arrive at the stable phase after progressing through all available metastable phases. Crystals may therefore nucleate in one metastable phase before transforming into another, thermodynamically more stable phase (Buanz et al.2019, Hilden et al. 2003, Lee et al. 2006).

### IV.1. Polymorphs and habits

Among the phases obtained with biomolecules, polymorphs have the same chemical composition but different structures giving different physicochemical properties, which can affect the manufacturing process, and the bioavailability (for instance for drugs) (Mangin et al. 2009, Lee 2014). It must be pointed out that different phases may have similar geometric appearance, which is called habit. Similarly, two crystals of the same phase may have different habits. Moreover, for statistical reasons droplet-based microfluidics is ideally suited to the detection of metastable phase, including new ones that have never been identified before in large volumes. Moreover, high values of $\beta$ accessible increases also the probability of nucleating a metastable phase (Mangin et al. 2009, Laval et al. 2008).

- For instance, cooling droplets of lysozyme generated with device of Fig. 16 leads to the appearance of one crystal in most of them. This mono-nucleation can be attributed to the depletion of lysozyme within the droplets as the first crystal grows, such that the supersaturation drops to a level that does not support a second nucleation event (Ildefonso et al. 2012b). A sea urchin–like habit different from the generally observed tetragonal crystal appeared in 6 droplets out of 237 (Fig. 27). In droplets of Fig. 27a (Ildefonso et al. 2012a), increasing the temperature to 30 °C dissolved only the crystal with the sea urchin–like habit (droplet on the right in Fig. 27b and c) and not the tetragonal crystal.

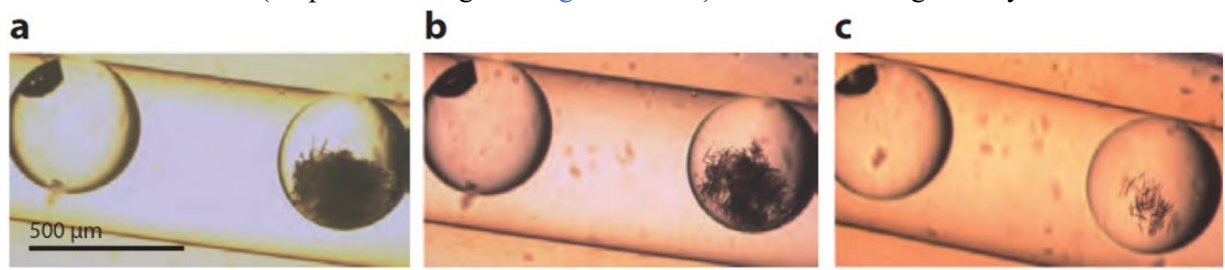

**Fig. 27** Images of 250 nL droplets of lysozyme solutions in device of figure 16: (*a*) after a storage of 20 h at 6 °C, (*b*) 6 min after increasing temperature to 30 °C, and (*c*) 12 min after increasing temperature to 30°C. Reprinted with permission from Ref. (Ildefonso et al. 2012a). Copyright (2012) Elsevier

Thus, we concluded that crystals in a sea urchin–like habit are a different phase (Astier and Veesler 2008). Crystals with this urchin-like habit were previously observed (Ataka and Asai 1988, Legrand et al. 2002) but were not easy to isolate in milliliter crystallizers. Moreover, owing to mononuclear nucleation, the nucleated crystal of the metastable phase cannot easily follow a solution-mediated phase transition to a more stable phase. Small volumes of droplets seem to freeze the metastable phase (Ildefonso et al. 2012a). It was also suggested that the reduction/elimination of convection in the droplets also reduces the rate of nucleation and growth, which retards the transformation of the metastable polymorph (Meldrum and O'Shaughness 2020). Thus, for statistical and kinetic reasons, microfluidics and small-volume, increase the chances of nucleating and stabilize metastable phase.

With phases of different compositions, such as solvates, the same phenomena were observed. For instance, we explored rasburicase with the device of Fig. 17 in a viscous media containing polyethylene glycol (PEG). The same experimental conditions produced different crystal habits and phases. Moreover, the two phases known for rasburicase (Vivares et al. 2006, Hu et al. 2010) appeared in different droplets with the same composition (Fig. 28) (Zhang et al. 2015b). The main reason for this concomitant nucleation is that the difference between their solubilities is small.

|   | Rasburicase | PEG | Temperature |
|---|---|---|---|
| a | 10 µg/µL | 10 % | 5° C |
| b | 10 µg/µL | 10 % | 20° C |
| c | 10 µg/µL | 5 % | 5° C |
| d | 5 µg/µL | 7,5 % | 5° C |
| e | 5 µg/µL | 7,5 % | 20° C |

**Fig. 28** Photos of crystals obtained after 24 h in 65 nL droplets of rasburicase, with a precipitant agent polyethylene glycol (PEG), generated with the device of Fig. 17: every line corresponds to the same condition. Reprinted with permission from Ref. (Zhang et al. 2015b). Copyright 2015 American Chemical Society

On the platform of Fig. 17, optical characterization of droplets enables counting the number of crystals in droplets, time of apparition, observation of habits and eventually phase transition. Crystal habits can indicate polymorphism; however, it is not sufficient evidence without further characterization.

### IV.2. Characterization of biomolecule crystals

The most widespread technique for the analysis of biomolecule crystals is XRD (see chapter 2 of part A of the present book), which provides information on the three-dimensional structure of crystals, and so on phases (see chapter 2 of part A of the present book). As authors (Table 1), we chose either to harvest the crystals of interest and make ex-situ XRD measurements or to do it in-situ.

*IV.2.1. Ex-situ XRD measurements*

Ex-situ XRD measurement was possible on our device of Fig. 17, because droplet containing crystal is in a Teflon tubing that could be cut. Then, the droplet was pushed out using a micro-injector and home-made micromanipulators for precise displacement in X, Y and Z (Grossier et al. 2010) and deposited on a MicroMesh grid in Kapton, which was previously treated in an oxygen plasma to make it hydrophilic. During deposition, the droplet spread over the MicroMesh grid, then detached, leaving only the crystal surrounded by the continuous phase FC70 oil (Fig. 29). The grid and crystal could then be removed from FC70 oil bath and immersed in liquid nitrogen to cryogenize the crystals. Here, FC70 oil acted as a cryoprotectant. However, crystals could also be immersed in a drop of glycerol for cryoprotection (Gerard et al. 2017).

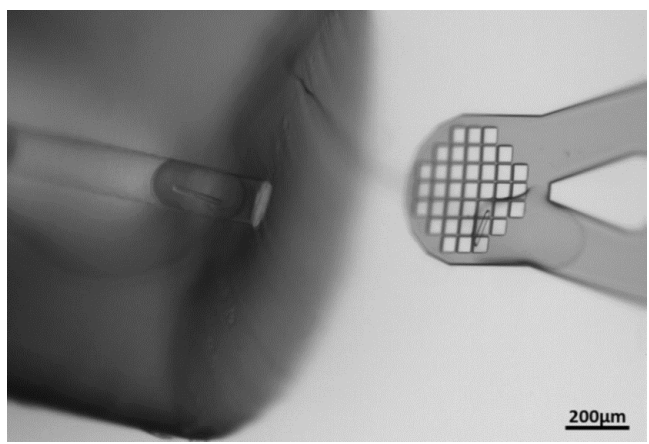

**Fig. 29** QR2 crystal nucleated in a Teflon tubing, deposited on the MicroMesh grid in a continuous phase bath. The drop spreads on the hydrophilic grid. Reprinted with permission from Ref. (Gerard et al. 2017). Copyright (2017) International Union of Crystallography

However, manual handling of the crystals can mechanically and environmentally stress them. The stress induced during this delicate step may affect the quality of the crystal and decrease its diffractive power. To minimize manual handling, an alternative is in situ X-ray data collection.

*IV.2.2. In-situ XRD measurements*

For in-situ XRD measurements, we transferred the droplets with crystal from Teflon tubing to silica tubing (fused silica tubing with a polyimide coating), using a linear junction. The internal silica tubing wall was coated with a commercial hydrophobic surface-coating agent (Fig. 30a), to ensure droplet stability. Then, the silica tubing containing the droplets was directly mounted on a magnetic base extracted from standard SPINE sample loops, ready for transfer to the X-ray setup (Fig. 30b). For a proof of concept, a single crystal of QR2 was analyzed by XRD at room temperature on the PROXIMA-1 beamline at Synchrotron SOLEIL and diffraction was observed to a resolution of 2.7 Å (Fig. 30c) (Gerard et al. 2017).

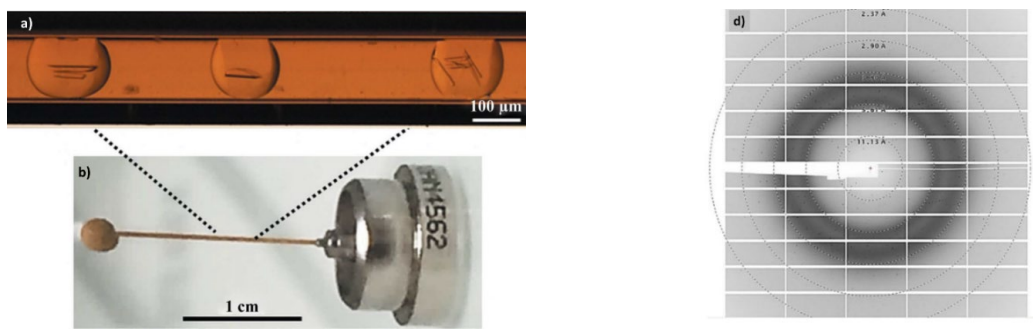

**Fig. 30** a) Silica tubing mounted on a b) magnetic base, with crystallized droplet of QR2 inside; c) Room-temperature in situ diffraction image of a QR2 crystal, with a resolution of 2.7 Å. Reprinted with permission from Ref. (Gerard et al. 2017). Copyright (2017) International Union of Crystallography

# Conclusion and Outlook

This chapter is dedicated to microfluidic devices used to study crystallization of biomolecules. We describe a variety of microfluidic devices developed to screen and optimize crystallization conditions of biomolecules, and to obtain crystals with controlled size, habit or phase. These microfluidic devices display micrometer-sized channels or wells (tens to few hundreds of micrometers), and thus crystallization proceeds in nanometric volumes (micro-crystallizers) that are used in the form of flows, droplets or wells/reservoirs. The methods of crystallization are inspired from vapor-diffusion, dialysis, counter-diffusion and batch methods that are typically used for biomolecule crystallization in large volumes. The particular case of droplet-based microfluidics answers the stochasticity of nucleation by enabling statistical approach: each droplet is an independent micro-crystallizer and a large number of experiments are carried out with hundreds of droplets under identical conditions.

By miniaturizing and parallelizing experiments, microfluidic devices enable playing on the quantity of precipitant and/or the temperature, and so screening different points of the phase diagram. The thermodynamic solubility curve and the metastable zone limit between nucleation and growth zones are determined. In droplet-based microfluidics, nucleation and crystal growth are decoupled to explore nucleation kinetic and nucleation frequency. It results in the determination of an effective interfacial energy crystal/solution, that reflects heterogeneous nucleation. Furthermore, the kinetic effect of confinement in nanovolumes generated in droplet-based microfluidics, leads to isolation of different phases that are characterized for structural application or identification of phases. Due to kinetic confinement, higher supersaturation is accessible in the solution before the solution nucleates, which increases the probability of nucleating a metastable phase. In addition, for statistical reasons droplet-based microfluidics is ideally suited to the detection of new metastable phases that have never been identified before in large volumes. Phases are identified by X-Ray Diffraction (XRD), either by harvesting the crystals of interest and make ex-situ measurements, or by in situ X-ray data collection on-line or off-line, to minimize manual handling.

For new users of microfluidic devices dedicated to biomolecule crystallization, it can be challenging to manage problems such as flows homogeneities, droplet size and concentration homogeneities, non-adhesion of crystals to channel walls... Moreover, biomolecules can exhibit effects on the interfacial stability and so on the rheological behaviours. These problems will be solved by playing on hydrodynamics, with improved structures and materials. Furthermore, the number of crystals in droplets remains challenging to predict, requiring the development of predictive models. Finally, analytical instruments associated to microfluidic devices should be accurate whatever the frequency of analysis. The future lies in the simplification and modularization of microfluidic devices, with units featuring basic functions for solution preparation, flow and/or droplet generation, incubation with temperature gradients, crystal detection and analysis. The idea could be to make all these functions compatible with any biomolecule and automatable. The, the challenge will be to scale up these microfluidic devices for industrial applications of biomolecules.

# NOMENCLATURE

$T$: temperatures (°C or K)

$T_{MZ}$: minimum temperature of the metastable zone (°C)

$T_{growth}$: temperature of growth (°C)

$T_{nucleation}$: temperature of nucleation (°C)

$\Delta t$: nucleation time (h)

$J$: nucleation rate (number of crystals·s$^{-1}$·m$^{-3}$)

$N$: average number of crystals nucleated in one droplet

$K_0$: kinetic factor (m$^{-3}$·s$^{-1}$)

$\Delta G^*_{HON}$: activation free energy for homogeneous nucleation (J)

$k$: Boltzmann constant (J·K$^{-1}$)

$\gamma$: isotropic interfacial energy of the critical nucleus (J·m$^{-2}$)

$\Omega$: volume of one molecule in the critical nucleus (m$^3$)

$\Delta G^*_{HEN}$: activation free energy required to form the critical cluster (J)

$\theta$: the contact angle of the cluster onto the foreign substance (deg)

$f(\theta)$: factor the thermodynamic part of the catalytic effect of the foreign substance on nucleation

$\gamma_{ef}$: an effective interfacial energy (J·m$^{-2}$)

$Cs$: thermodynamic solubility (g/L)

$C$: concentration (g/L)

$\beta$: supersaturation

$t_N$: average nucleation induction time for a single nucleus (h)

$V$: volume of the crystallizer (m$^3$)

## List of Abbreviations

XRD : X-Ray Diffraction

PDMS: polydimethylsiloxane

COC: Cyclic Olefin Copolymer

PMMA: Polymethylmethacrylate

PEEK: PolyEther Eher Ketone

PFA: Poly Fluoropolymerssuch as Perfluoroalkoxy

FEP: Fluorinated Ethylene Propylene

PTFE: Polytetrafluoroethylene

THV: Tetrafluoroethylene, hexafluoropropylene and vinylidene fluoride

PP: Polypropylene

FMS: Fluoropropylmethylsiloxane

PEG: polyethylene glycol

BIM: Barrier Interface Metering

FID: Free-interface diffusion

CaMKIIβ: Calcium–calmodulin dependent kinase II

CNT: Classical Nucleation Theory

RC: Reaction center

Rel.: Relative

QR2 : Human quinone reductase 2

MZ: Metastable Zone

MZL: Metastable zone limit

DPT: double-pulse technique

HON: Homogeneous Nucleation

HEN: Heterogeneous Nucleation